\input epsf
\documentclass[nohyper,notoc]{article} 
\usepackage{axodraw}
\usepackage{epsfig}
\usepackage{bm}
\usepackage{color}

\usepackage[utf8]{inputenc} 
\usepackage[T1]{fontenc}    

\def\ben{\begin{enumerate}}
\def\een{\end{enumerate}}
\def\bit{\begin{itemize}}
\def\eit{\end{itemize}}
\def\beq{\begin{equation}}
\def\eeq{\end{equation}}
\def\bea{\begin{eqnarray}}
\def\eea{\end{eqnarray}}
\def\bq{\begin{quote}}
\def\eq{\end{quote}}
\def \lsim{\mathrel{\vcenter
     {\hbox{$<$}\nointerlineskip\hbox{$\sim$}}}}
\def \gsim{\mathrel{\vcenter
     {\hbox{$>$}\nointerlineskip\hbox{$\sim$}}}}
\def\gappeq{\mathrel{\rlap {\raise.5ex\hbox{$>$}}
{\lower.5ex\hbox{$\sim$}}}}
\def\lappeq{\mathrel{\rlap{\raise.5ex\hbox{$<$}}
{\lower.5ex\hbox{$\sim$}}}}

\def\meg{\mu \to e \gamma}

\def\a{\alpha}
\def\b{\beta}
\def\g{\gamma}

\def\Lamf{\Lambda_6}

\def\ifmath#1{\relax\ifmmode #1\else $#1$\fi}
\def\lsim{\mathrel{\raise.3ex\hbox{$<$\kern-.75em\lower1ex\hbox{$\sim$}}}}
\def\gsim{\mathrel{\raise.3ex\hbox{$>$\kern-.75em\lower1ex\hbox{$\sim$}}}}

\def\eq#1{eq.~(\ref{#1})}

\def\anti{\overline}

\def\rhu{\rho^U}

\def\rhd{\rho^D}
\def\rhe{\rho^E}

\def\cbma{c_{\beta-\alpha}}

\def\sbma{s_{\beta-\alpha}}

\def\ifmath#1{\relax\ifmmode #1\else $#1$\fi}

\def\call{\mathcal{L}}
\def\s2W{\sin^2\theta_W}

\def\tanb{\tan\beta}
\def\cotb{\cot\beta}

\def\hl{h}
\def\ha{A}
\def\hh{H}

\def\mhl{m_{\hl}}
\def\mhh{m_{\hh}}

\def\ls#1{\ifmath{_{\lower1.5pt\hbox{$\scriptstyle #1$}}}}
\def\lss#1{\ifmath{^{\,\lower2.5pt\hbox{$\scriptstyle #1$}}}}

\def\half{\ifmath{{\textstyle{1 \over 2}}}}

\begin{document}

\renewcommand{\thefootnote}{\fnsymbol{footnote}}
\begin{center}
{\Large {\bf
$\meg$ in the 2HDM: an exercise in EFT}
}
\vskip 25pt
{\bf   Sacha Davidson \footnote{E-mail address:
s.davidson@ipnl.in2p3.fr}  
}
 
\vskip 10pt  
{\it IPNL, CNRS/IN2P3,  4 rue E. Fermi, 69622 Villeurbanne cedex, France; 
Universit\'e Lyon 1, Villeurbanne;
 Universit\'e de Lyon, F-69622, Lyon, France
}\\
\vskip 20pt
{\bf Abstract}
\end{center}
\begin{quotation}
{\small 
\noindent
The 2 Higgs Doublet Model (2HDM) of type III has renormalisable Lepton
Flavour-Violating couplings, and its one and two-loop (``Barr-Zee'')  
contributions to   $\meg$ are known. 
In the decoupling limit, where  the  mass
scale $M$ of the second  doublet is much greater
than the electroweak scale, the model can be parametrised
with an  Effective Field Theory(EFT) containing
dimension six operators.
The $1/M^2$ terms of the exact calculation are
reproduced in the EFT, provided that  
the four-fermion operator basis   below  the weak scale
is enlarged  with respect to the SU(2)-invariant Buchmuller-Wyler list. 
It is found that the dominant two-loop ``Barr-Zee'' contributions 
arise mostly in 
two-loop matching and running,  
and  that dimension eight operators  can be numerically relevant.

}
\end{quotation}

\vskip 20pt  

\setcounter{footnote}{0}
\renewcommand{\thefootnote}{\arabic{footnote}}

\section{Introduction}
\label{intro}

This exercise was born from a puzzle:
 experiments 
that search for $\mu \leftrightarrow e$ flavour change
constrain a
long list of   QCD$\times$QED invariant four-fermion operators,
some of which   turn out to be 
  dimension eight  when  SU(2) invariance is imposed.
But it is common, when   describing New Physics
from above $m_W$ with  Effective
Field Theory(EFT)\cite{Georgi},   to
use the  SU(2) invariant  basis of dimension 6 operators
given by Buchmuller and Wyler\cite{BW} and pruned in \cite{Polonais}.
To explore when it is justified to neglect
the additional four-fermion operators
below $m_W$, 
 we were looking for a model where 
they  might  give  relevant contributions. The first
model
we tried was the 2 Higgs Doublet Model (2HDM).
It turns out  that
the  additional  four-fermion operators 
(not in the Buchmuller-Wyler list)
must be included
below $m_W$ to correctly reproduce the ${\cal O}(1/M^2)$
terms in the $\meg$ amplitude of the 2HDM.

 The  exercise takes place in
a  Type III 2HDM in the decoupling limit, where
the one-loop and two-loop ``Barr-Zee''
contributions to  the $\meg$  amplitude are known\cite{CHK}.
The aim is  to to  extract  the numerically
dominant contributions,  and identify where they arise in 
an EFT description.

Section \ref{sec:2HDM} reviews the 2HDM of Type III in
the decoupling limit, and the 
calculation in this model   of  the $\meg$ amplitude by
Chang, Keung and Hou \cite{CHK} (CHK). 
Type III 2HDMs include  charged-lepton 
flavour-changing couplings; for simplicity, 
only a $\mu \leftrightarrow e$ flavour-changing interaction is allowed.
The decoupling limit is taken, by
requiring the mass scale $M$  of the  second doublet 
 to be $\gsim  10 v$, to ensure that EFT 
can give a reasonable approximation to
the $\meg$ amplitude in this model.
In the final subsections, the 
${\cal O}(1/M^2)$ and ${\cal O}(1/M^4)$
parts of the $\meg$ amplitude
are compared, for the various classes of
diagrams.

Section \ref{sec:EFT} sets up an EFT formalism,
based on dimension six operators and
one-loop RGEs, which correctly  reproduces
all the ${\cal O}([\alpha \log]^n/M^2)$ terms of
the CHK calculation in the full model.  How to obtain the remaining
terms of the CHK result is  outlined
in Appendix \ref{ssec:dim8}.

Section \ref{sec:disc} summarises
the less trivial aspects of the calculation, and 
the location in the EFT of the Barr-Zee diagrams.
It should be readable
independently of the more technical sections
\ref{sec:2HDM} and \ref{sec:EFT}.
 There are two-and-a-half issues with the results
obtained with dimension six operators and one-loop RGEs: the  four-fermion 
operator basis below $m_W$ only needs to
be invariant under QCD and QED, so it should be enlarged
with respect to the  list\cite{Polonais} of
SU(2)-invariant  dimension six operators.
Second, two-loop effects are important, because the one-loop contribution is
suppressed by  the square of the muon Yukawa coupling.
Finally, dimension eight terms
can be enhanced, both by
logs and by  unknown couplings  in the 2HDM
(encapsulated in $\tan \beta$).
 
This exercise overlaps with several studies.
Pruna and Signer\cite{PS}  studied $\meg$ in an EFT
consisting of the SM extended by  a
complete set of SU(2)-invariant operators,
but they focused on the electroweak running
above $m_W$,  rather than the matching
at $m_W$, so did not extend the operator
basis below $m_W$. In the context
of $B$ physics, Alonso, Grinstein and
Camalich \cite{AGC}, and
Aebischer etal\cite{ACFG} calculated  
the coefficients of  the enlarged operator
basis below $m_W$, given a selection
of SU(2) invariant operators above $m_W$.
This exercise only agrees approximatively
with  \cite{AGC}, as discussed in section \ref{sec:disc}.


\section{$\meg$ in the type III 2HDM}
\label{sec:2HDM}

\subsection{Review of the 2HDM type III in the decoupling limit}

The 2HDM is a minimal extension of the Standard Model, including one
extra Higgs doublet  with new unknown interactions to the
Standard Model fermions and Higgs --- for a review,
see \cite{2HDMrev}.  
In the 2HDM considered here, 
the extra  doublet is taken to be heavy --- this is the decoupling
limit, and should be describable  with EFT. To allow for LFV,
consider a ``type III'' model, where there is no discrete
symmetry  that  distinguishes the Higgs,  so no
``symmetry basis'' in which to write the Lagrangian
(so also no  unambiguous 
definition of $\tan \beta$). For simplicity,
the Higgs potential is taken CP invariant.

The Lagrangian can be written in the ``Higgs basis'', 
 defined such that  
$\langle H_1 \rangle \neq 0$, 
and  $\langle H_2\rangle =0$,  so the doublets  are written
\footnote{Neglecting the phase ambiguity $\chi$ discussed at
eqn A.11 of \cite{Howie}}
\beq \label{abbasis}
H_1=\left(\begin{array}{c}
G^+ \\ {1\over\sqrt{2}}\left(v+H_1^0+iG^0\right)\end{array}
\right)\,,\qquad
H_2=\left(\begin{array}{c}
H^+ \\ {1\over\sqrt{2}}\left(H_2^0+i\ha\right)\end{array}
\right)\,,
\eeq
 where
the $G$s are goldstones. This shows that  the mass eigenstates
$A$ and $H^\pm$ are the
CP-odd and charged components of the
vev-less $H_2$, but there can be some mixing
between $H_1$ and $H_2$ in the CP-even scalars $h,H$.
The convention $\langle H_1 \rangle = v/\sqrt{2}$
implies $v= 246$ GeV, 
\beq
\frac{4 G_F}{\sqrt{2}} = \frac{2}{v^2}
~~~,~~~\sqrt{2} G_F = \frac{1}{v^2}~~~.
\eeq
In this ``Higgs basis'',  in the notation
of \cite{Howie},  the 
potential parameters are written in upper case, so the
potential is:
\bea
\mathcal{V}&=& M_{11}^2H_1^\dagger H_1+M^2 H_2^\dagger H_2
-[M_{12}^2 H_1^\dagger H_2+{\rm h.c.}]~~~~~~~~~~~~~~ {\rm (in~ Higgs~ basis)}
\nonumber\\[6pt]
&&\quad +\half\Lambda_1(H_1^\dagger H_1)^2
+\half\Lambda_2(H_2^\dagger H_2)^2
+\Lambda_3(H_1^\dagger H_1)(H_2^\dagger H_2)
+\Lambda_4(H_1^\dagger H_2)(H_2^\dagger H_1)
\nonumber\\[6pt]
&&\quad +\left\{\half\Lambda_5(H_1^\dagger H_2)^2
+\big[\Lambda_6(H_1^\dagger H_1)
+\Lambda_7(H_2^\dagger H_2)\big]
H_1^\dagger H_2+{\rm h.c.}\right\}\,,
\label{V}
\eea

The angle $\beta - \alpha$
can  be  defined from the Higgs potential of type a III model,
unlike $\beta$ and $\alpha$.
It rotates between
the mass basis of  $h,H$ 
and the Higgs basis:
\bea
\hl &=& H_1^0~\sbma+ H_2^0\,\cbma\,,\nonumber\\
\hh &=& H_1^0~\cbma-
H_2^0\,\sbma\,,
\label{hscalareigenstates}
\eea
so that what is here called
 $\beta -\alpha$ is independent of the
angle  $\beta$ that will later be  defined from the Yukawas.
If the potential is CP-invariant, there is a
simple relation for $\cbma$  \cite{Howie}:
\bea \label{exactbma}
\cos\,[(\beta-\alpha)] \sin\,[(\beta-\alpha)]&=&{-\Lamf v^2\over\mhh^2-\mhl^2}\,
\eea

The  masses of the scalars are related to potential parameters
in the Higgs basis as \cite{Howie} :
\bea
m_{H^\pm}^2 & = & M^2 + \frac{v^2}{2} \Lambda_3 \nonumber \\
m_A^2 - m_{H^\pm}^2&= & -\frac{v^2}{2} (\Lambda_5 - \Lambda_4) \nonumber \\
m_H^2 + m_h^2-  m_A^2 & = & + v^2 (\Lambda_1 + \Lambda_5)  \nonumber\\ 
(m_H^2 - m_h^2)^2 & = & 
[m_A^2 + (\Lambda_5 - \Lambda_1)v^2]^2 + 4 \Lambda_6^2 v^4 
\label{exactmasses}
\eea
and the couplings to  $W^+W^-$ are 
 $ i g m_W C_{\phi WW} 
 g^{\mu \nu}$
with
\beq
 C_{h WW} = s_{ \b - \a} ~,~ C_{H WW} = c_{\b - \a} ~,~  
C_{A WW} =0
\label{phiWW}
\eeq

In the decoupling limit \cite{GunionHaber}  where $\Lambda_i v^2 \ll M^2$,
 the exact relations (\ref{exactmasses}) can be
expanded  in $v^2/M^2$  to obtain:
\bea
m_H^2 - m_{A}^2& \simeq  & +\Lambda_5 v^2~~~~~~~~~~~~~~~~~~~~~~~~~~~~~~~~~~~ ~~~~~~~~({\rm decoupling~limit}) \nonumber \\
m_h^2  & \simeq & + v^2 \Lambda_1  - \frac{v^4}{M^2} 
\left(\Lambda_6 +
 \frac{(\Lambda_5 - \Lambda_1)^2}{4} \right) ~~~~~~~~({\rm decoupling~limit})
\label{eq13}
\eea
and then   approximating $\sbma \simeq 1 $ in eqn
(\ref{exactbma}), gives
\bea
\Rightarrow \cos\,[(\beta-\alpha)] \simeq {-\Lamf v^2\over M^2}
\left(1 + \frac{\Lambda_1v^2}{M^2} \right) + ...
~~~~~~~~({\rm decoupling~limit})
\label{cbmadl}
\eea
which confirms that in
decoupling limit,
$h$ is mostly  $H_1$, and $H$ is mostly $H_2$.

The Yukawa couplings in
  the Higgs basis for the Higgses, and the mass
eigenstate basis for the $\{ u_R, d_R,e_R, d_L, e_L \}$, are:
\bea \label{ymodel2}
\!\!\!\!\!\!\!\!
-{\cal L}_{\rm Y}&=&   {\Big (}
\overline{Q}_{j}  \widetilde{H}_1 K^*_{ij} Y^{U}_i U_{i}   +
\overline{Q}_i H_1  Y^{D}_{i}  D_{i}
+\overline{L}_i H_1  Y^{E}_{i}  E_{i} {\Big ) }
\nonumber \\
&&
+\overline{Q}_{i}  \widetilde{H}_2 [K^\dagger \rho^{U}]_{ij} U_{j}   +
\overline{Q}_i H_2 [\rho^{D}]_{ij} D_{j}
+\overline{L}_i H_2 [\rho^{E}]_{ij} E_{j}
+{\rm h.c.}\,,
\eea
where $Q,L$ are SU(2) doublets, $E,U,D$ are singlets,
  SU(2) indices are implicit
($\overline {L} H_1 = \bar{\nu} H^+_1 + \bar{e} H^0_1$),
 $\widetilde{H}_i = i \sigma_2  H_i^*$,  
 the generation indices are explicit,
 and $K$ is the CKM matrix. 
The $Y$ matrices
are flavour diagonal  and equal to the SM Yukawas:
\beq
[Y^P]_{ij} = \sqrt{2}  \frac{m^P_j}{ v} \delta_{ij}
\eeq
as a result of being in the fermion mass eigenstate bases.

The $\rho^P$ matrices  can be flavour-changing. To obtain
$\meg$ without other flavour-changing processes, I
neglect all  off-diagonal elements except 
$\rho_{\mu e} \simeq \rho_{ e \mu} \neq 0$.
To obtain a predictive model, and make contact with
other 2HDM literature, the diagonal elements 
follow the pattern of one  of the types of
 2HDM which have a discrete symmetry that
 ensures flavour conservation:
\beq
\begin{array}{c|ccc}
{\rm Model ~type} &\rho^U  & \rho^D  & \rho^E  \\
\hline
{\rm Type ~I}  &Y^U \cotb & - Y^D\tanb & -Y^E \tanb\\
{\rm Type ~II} &-Y^U \tanb & - Y^D\tanb & -Y^E \tanb \\
{\rm Type~ X  }&Y^U \cotb &  Y^D\cotb & -Y^E \tanb \\
{\rm Type~ Y} &-Y^U \tanb &  Y^D\cotb &  -Y^E \tanb \\
\end{array}
 \label{ktb}
\eeq
This requires a definition of $\tanb$ which
is common to all the fermions. It can
for instance be defined   from the  $\tau$ Yukawa:
in the mass eigenstate basis  for $\tau_R, \tau_L$, define
$H_\tau$ to be the linear combination of Higgses
to which couples the $\tau$,  and  $\b$ as the
angle in Higgs doublet space 
between  $H_1$ (the vev) and $H_\tau$:
\bea
H_{\perp} &  = & \widetilde{H}_1 \sin \b +
\widetilde{H}_2 \cos \b \nonumber  \\ 
H_\tau  &  = & {H}_1 \cos \b - 
{H}_2 \sin \b
\label{tanb}
\eea 
Since $\tan \beta$ is defined from the lepton Yukawas,
$\rho^E \propto \tan \beta$ in all Types
of 2HDM listed above. This is unconventional,
but should include the same predictions provided
that $\tan \beta$ is allowed to range from
$1/50 \to 50$.


From eqn (\ref{ymodel2}) the  couplings to fermions  are 
\bea \label{modeliiihqq}
-\call_Y&=&
\anti d \frac{ 1  }{\sqrt{2}}
 \left[ Y^D  (P_R + P_L) \sbma+
(\rhd P_R+{\rhd}^\dagger P_L)\cbma\right]d  \hl   \nonumber \\[4pt]
&&+ \anti d \frac{ 1  }{\sqrt{2}}
\left[ Y^D(P_R + P_L) \cbma-
(\rhd P_R+{\rhd}^\dagger P_L)\sbma\right]d\hh
+\frac{i}{\sqrt{2}}\anti d(\rhd P_R-{\rhd}^\dagger P_L)d
\ha\nonumber\\[4pt]
&&\anti e \frac{ 1  }{\sqrt{2}} 
\left[ Y^E(P_R + P_L) \sbma+
(\rhe P_R+{\rhe}^\dagger P_L)\cbma\right]e\hl  \nonumber \\[4pt]
&&+ \anti e 
\frac{ 1  }{\sqrt{2}}
\left[ Y^E(P_R + P_L) \cbma-
(\rhe P_R+{\rhe}^\dagger P_L)\sbma\right]e \hh
+\frac{i}{\sqrt{2}}\anti e(\rhe P_R-{\rhe}^\dagger P_L)e
\ha\nonumber\\[4pt]
&&+ \anti u
\frac{ 1  }{\sqrt{2}}
\left[Y^U(P_R + P_L) \sbma+
( \rhu P_R+{\rhu}^\dagger  P_L)\cbma\right]u\hl
 \nonumber \\[4pt]
&&+\anti u \frac{ 1  }{\sqrt{2}}
\left[ Y^U (P_R + P_L) \cbma-
(\rhu P_R+{ \rhu}^\dagger P_L)\sbma\right]u\hh
-\frac{i}{\sqrt{2}}\anti u( \rhu P_R-{\rhu}^\dagger P_L)u\ha 
\nonumber \\[4pt]
&&+\left\{\anti u\left[K\rhd P_R-{\rhu}^\dagger KP_L\right] dH^+
 +{\rm h.c.}\right\}
+\left\{\anti \nu \left[\rhe P_R \right] e H^+
 +{\rm h.c.}\right\}\,.
\eea
This gives Feynman rules of the form $-iF^{\phi,X}_{ij} P_X$, where:
\bea  \label{modeliiicouplings}
F^{\hl,L}_ {ij}=
\frac{Y^P_{ij}}{\sqrt{2}}\sbma+ \frac{[\rho^{P\dagger}]_{ij}}{\sqrt{2}}\cbma ~~~&,&
~~~
F^{\hl,R}_ {ij}=
\frac{Y^P_{ij}}{\sqrt{2}} \sbma+ \frac{[\rho^{P }]_{ij}}{\sqrt{2}}\cbma
\\ \nonumber
F^{\hh,L}_ {ij}=
\frac{Y^P_{ij}}{\sqrt{2}}\cbma -\frac{[\rho^{P \dagger}]_{ij}}{\sqrt{2}}\sbma ~~~&,&
~~~
F^{\hh,R}_ {ij}=
\frac{Y^P_{ij}}{\sqrt{2}} \cbma - \frac{[\rho^{P }]_{ij}}{\sqrt{2}}\sbma
\\
F^{\ha,L}_ {u_iu_j}=
 i\frac{[\rho^{U \dagger}]_{ij}}{\sqrt{2}} ~~~&,&
~~~
F^{\ha,R}_ {u_iu_j}=
-i \frac{[\rho^{U }]_{ij}}{\sqrt{2}}
\nonumber \\
F^{\ha,L}_ {i j}=
- i\frac{[\rho^{P \dagger}]_{ij}}{\sqrt{2}} ~~~&,&
~~~
F^{\ha,R}_ {ij}=
i \frac{[\rho^{P }]_{ij}}{\sqrt{2}}
~~~~,{\rm for} ~P \in ~E,D~~,
\nonumber
\eea
where the possible forms for  $\rho^P$  are given in eqn \ref{ktb}.

\subsection{$\meg$ in the 2HDM}
\label{}

The decay $\meg$ can be parametrised by adding 
 the dipole operator to the 
Standard Model Lagrangian. In the
notation of Kuno  and Okada\cite{KunoOkada}
\beq
{\cal L}_{meg} = -\frac{4G_F}{\sqrt{2}} m_\mu \left(
A_R \overline{\mu_R} \sigma^{\a\b} e_L F_{\a\b} +
A_L \overline{\mu_L} \sigma^{\a\b} e_R F_{\a\b}\right)
\label{Lmeg}
\eeq
which gives
\beq
BR(\meg)= 384 \pi^2 (|A_R|^2 + |A_L|^2) <5.7\times 10^{-13}
\label{BRmeg}
\eeq
where the  upper bound is from the MEG experiment \cite{MEG13}.
(The amplification factor of $384\pi^2 = \frac{3}{2}(16 \pi)^2$  is because
$\meg$ is a 2-body decay, and is compared to  the
usual three-body muon  decay.).  $|A_R|, |A_L|$ are
dimensionless, and
if $|A_R| = |A_L|$, then $|A_X| <8.6\times 10^{-9}$.

\begin{figure}[ht]
\unitlength.5mm
\SetScale{1.418}
\begin{boldmath}
\begin{center}
\begin{picture}(60,80)(0,0)
\ArrowLine(0,20)(20,20)
\ArrowLine(20,20)(60,20)
\ArrowLine(60,20)(80,20)
\DashCArc(40,20)(20,0,180){1}
\Text(-2,13)[r]{$\mu$}
\Text(82,13)[l]{$e$}
\Text(45,-10)[l]{$\gamma$}
\Text(63,35)[r]{$\phi$}
\Text(20,15)[c]{$F^\phi_{\mu \mu}$}
\Text(35,27)[c]{$\mu$}
\Text(49,20)[c]{${\rm x}$}
\Photon(40,20)(40,-10){2}{4}
\Text(60,15)[c]{$F^\phi_{e \mu}$}
\end{picture}
\qquad\qquad\qquad
\begin{picture}(60,80)(0,0)
\ArrowLine(0,0)(20,0)
\ArrowLine(20,0)(60,0)
\ArrowLine(60,0)(80,0)
\DashLine(60,0)(48,18){1}
\CArc(40,25)(10,0,360)
\Photon(40,35)(40,60){2}{4}
\Photon(33,18)(20,0){2}{4}
\Text(10,-5)[r]{$\mu$}
\Text(72,-5)[l]{$e$}
\Text(48,50)[l]{$\gamma$}
\Text(52,35)[l]{$t, b$}
\Text(16,15)[c]{$\gamma,Z$}
\Text(60,15)[r]{$\phi$}
\Text(60,-5)[c]{$F^\phi_{e \mu }$}
\end{picture}
\qquad\qquad\qquad
%
\begin{picture}(60,80)(0,0)
\ArrowLine(0,0)(20,0)
\ArrowLine(20,0)(60,0)
\ArrowLine(60,0)(80,0)
\DashLine(60,0)(48,18){1}
\PhotonArc(40,25)(10,0,360){2}{7}
\Photon(40,35)(40,60){2}{4}
\Photon(33,18)(20,0){2}{4}
\Text(-2,-5)[r]{$\mu$}
\Text(72,-5)[l]{$e$}
\Text(48,50)[l]{$\gamma$}
\Text(57,35)[r]{$W$}
\Text(16,15)[c]{$\gamma,Z$}
\Text(60,15)[r]{$\phi$}
\Text(60,-5)[c]{$F^\phi_{e \mu }$}
\end{picture}
\end{center}
\end{boldmath}
\vspace{10mm}
\caption{One and two-loop diagrams contributing to $\meg$ in the 2HDM, 
in the presence of a flavour-changing Yukawa coupling $F_{\mu e}$(see
eqns (\ref{modeliiicouplings})).
\label{fig:1loop}}
\end{figure}
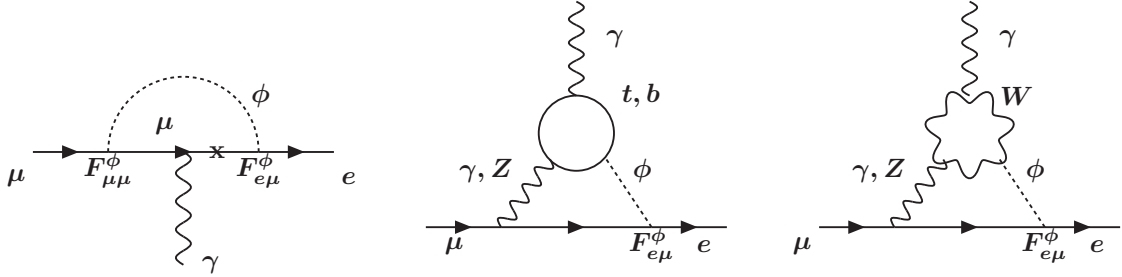

The decay $\meg$ has been extensively studied in the 
2HDM \cite{BW,hLFV,Grenier}, particularily in connection
\cite{hLFVrecent}
with the recent LHC\cite{hLFVCMS,OST} excess in $h\to \tau^\pm \mu^\mp$.
Chang,Hou and Keung \cite{CHK}(CHK) 
calculate  the  contributions to $\meg$  of
neutral Higgs bosons with flavour-changing couplings.
Their calculation can be divided into four
classes of diagrams, illustrated in figure \ref{fig:1loop}:
the one-loop diagrams, then three classes of two loop diagrams,
that is, those
with a $t$ loop, with a $b$ loop  and with a $W$ loop.  
I use their results, and later,  in matching onto
SU(2) invariant operators, assume that the charged Higgs
contribution ensures the SU(2) invariance of the
results. The results of CHK are refered to as
``full-model'' results in the following.

In the appendix is given a translation dictionary between
the notation of CHK  and here.   The amplitude 
given by CHK \cite{CHK}\footnote{The constant  factors
given  by Omura {\it et.al.} \cite{OST} differ:-$\frac{4}{3}$ for $h,H$ and
-$\frac{5}{3}$ for $A$. However, the constant
is irrelevant here because the  aim here is only
to reproduce the log in EFT. Also as noted  by Omura  {\it et.al.},
there are doubtful signs
and typos/missing terms in the Barr-Zee formulae in
 \cite{Grenier}, which differ from the formulae here. }
for the  one-loop diagram of figure \ref{fig:1loop},
with an internal  $\mu$, is
\bea
-2\sqrt{2} G_F m_\mu A^{1~loop}_{L}& = &-\frac{em_\mu}{32 \pi^2} 
\left[\sum_{\phi = h,H,A} \frac{ F^{\phi,L}_{\mu \mu} F^{\phi,L}_{e \mu } }{m_\phi^2} 
\left(\ln \frac{m_\mu^2}{m_\phi^2} + \frac{3}{2} \right) \right]^*
\label{1lexact}
\\
-2\sqrt{2} G_F m_\mu A^{1~loop}_{R} & = &-\frac{e m_\mu}{32 \pi^2} 
\left[\sum_{\phi = h,H,A} \frac{ F^{\phi,R}_{\mu \mu} F^{\phi,R}_{e \mu } }{m_\phi^2} 
\left(\ln \frac{m_\mu^2}{m_\phi^2} + \frac{3}{2} \right) \right]^*
\nonumber
\eea
Notice that  these amplitudes are suppressed by two
leptonic Yukawas and an additional muon mass
insertion to flip the chirality. 
In  the decoupling limit,
with $\phi \in \{H,A\}$ and $\mu$ in the loop, eqn(\ref{1lexact}) for $A_L$
gives
\bea
-2\sqrt{2} G_F m_\mu A^{H,A,1~loop}_{L}&  \simeq & 
-\frac{e m_\mu}{32 \pi^2M^2} [\rho^E]_{\mu e} 
\left[   Y^E_{\mu \mu} \frac{ \Lambda_6 v^2}{2 M^2}  
- \frac{[\rho^E]_{\mu \mu} }{2M^2} \Lambda_5v^2
 \right]^*  \ln \frac{m_\mu^2}{M^2}
\label{1ldc}
\eea
where  $m_H \simeq m_A \simeq M$ was used  in the logarithm,
and  terms suppressed by more than  $M^{-4}$ were dropped.
 The result for $A_L$   is  obtained by replacing 
$[\rho^E]_{ij} \to [\rho^E]^*_{ij }$ in the above. The one-loop diagram with
the light Higgs $h$ and a $\mu$ (also in the the decoupling limit),
 gives 
\bea
-2\sqrt{2} G_F m_\mu A^{h,1~loop}_{L}&   \simeq &
-\frac{e m_\mu}{32 \pi^2m_h^2} 
[\rho^E]_{\mu e}
\left[ -Y^E_{\mu \mu} \frac{ \Lambda_6 v^2}{2 M^2}
 + 2 
[\rho^E]_{\mu \mu}  \left( \frac{ \Lambda_6 v^2}{2 M^2}\right)^2
\right]^*
 \ln \frac{m_\mu^2}{m_h^2}  ~~~.
\label{1lh}
\eea

As noted long ago by Bjorken and Weinberg\cite{bjw}, 
there are   two-loop  diagrams  which 
 can be relevant  for $\meg$. Some examples
are illustrated to the right in figure \ref{fig:1loop};
a more complete set of diagrams with broken
electroweak symmetry can be found in \cite{LPX}. 
The resulting two-loop  contributions
to $\meg$  can be  numerically larger than the one-loop
contributions,   because the
Higgs attaches only once to the lepton line,
via the flavour-changing coupling, and 
otherwise couples to a $W,b$ or $t$ loop.
The result for the top loop (neglecting 
the diagrams with  internal $Z$
exchange, see figure \ref{fig:1loop}), is
\bea
-2\sqrt{2} G_F m_\mu A_{L}^{t~loop} &=&\frac{e\alpha }{16\pi^3}  
\frac{1}{m_t} 3Q_t^2
\left[ \sum_{\phi = h,H} F_{e \mu}^{\phi,L}  F^{\phi,L}_{tt} 
f(\frac{m_t^2}{m_\phi^2}) + 
F_{e \mu}^{A,L}   F^{A,L}_{tt}  
g(\frac{m_t^2}{m_A^2}) \right]^*
\nonumber \\
 & \simeq &\frac{e\alpha m_t }{32\pi^3M^2}   [\rho^{E }]_{\mu e}  3Q_t^2
\left[    \left( [\rho^{U}]_{tt} +  Y^{U}_{tt}\frac{\Lambda_6 v^2}{2M^2}  
\right) \ln^2 \frac{m_t^2}{M^2}  
 -  
 2 Y^{U}_{tt} \Lambda_6 
f(\frac{m_t^2}{m_h^2}) \right]
\label{fonctionnett}
\eea
where the log$^2$ terms  of the  approximate equality 
   are the sum of the heavy $H$ and $A$ contributions,
and the third  term is  due to the light $h$.
The approximate equality is in the decoupling limit,
so uses $m_H \simeq m_A \simeq M$ in the log, 
neglects terms suppressed by $v^2/M^2$,
and  uses that, for small $z$ \cite{CHK},   
\beq
f(z) \simeq g(z) \simeq \frac{z}{2} \ln^2 z  ~~,~~
h(z) \simeq z \ln z
~~,~~ f(z) - g(z) \simeq z (\ln z +2)
~~.
\label{idfg}
\eeq
(The function $h$ will appear in the $W$ loop contribution.)
The functions $f,g$ and $h$ are given in CHK,  are slowly
varying near $z\sim 1$ and at $z = m_W^2/m_h^2\simeq 0.4$ they are $f \sim 0.7,
g\sim 0.9$ and $h\sim 0.5$.
The formula for $A^{t~loop}_R$  is obtained by  replacing 
$ F_{ij}^{H,L} \to  F_{ij}^{H,R}$ in the coefficient
of $f$, and $\rho^{F \dagger} \to \rho^{F}$
in the coefficient of $g$.
The top-loop amplitude is $\propto m_t$, as expected
because  a top mass insertion is required
both to have an even number of $\gamma$-matrices
in the loop, and to provide the Higgs leg of the 
dipole operator. 

A $b$-quark loop  should be described
by the same formula as the top loop
(again neglecting the $Z$-exchange diagrams),
  but  the $A$-exchange contribution will subtract
from $H$-exchange, because of
the  sign difference in the couplings
of $A$ to $b$s and $t$s (see eqn (\ref{modeliiicouplings})):
\bea
-2 \sqrt{2} G_F m_\mu A_{L}^{b~loop} 
&=&\frac{e\alpha  }{16 \pi^3 }  \frac{3Q_b^2}{m_b } 
\left[\sum_{\phi = h,H} F_{e \mu}^{\phi,L}  F^{\phi,L}_{bb} 
f(\frac{m_b^2}{m_\phi^2}) +
 F_{e \mu}^{A,L} F^{A,L}_{bb} 
g(\frac{m_b^2}{m_A^2}) \right]^*
\nonumber \\
 &\simeq &\frac{e\alpha m_b }{64\pi^3M^2}    3Q_b^2
     \rho^{E }_{ \mu e}  \left[ \frac{ v^2}{M^2} \left( Y^{D}_{bb} \Lambda_6   
- \rho^{D}_{bb} \Lambda_5 
\right)\ln^2 \frac{m_b^2}{M^2}
- \frac{v^2}{m_h^2} \left(  Y^{D}_{bb} \Lambda_6
-
\rho^{D }_{bb} \frac{\Lambda^2_6 v^2}{M^2}\right) \ln^2 (\frac{m_b^2}{m_h^2})  
\right]~~~~~
\label{fonctionnebb}
\eea
where the approximate equality is in the decoupling limit,
and uses the approximations eqn (\ref{idfg}).
The last term is the light higgs contribution.
Notice that the  ${\cal O} (1/M^2)$ contributions of $H$
and $A$ cancelled against each other in the decoupling limit.
The formula for $A_R$  is obtained by  replacing 
$ F_{ij}^{H,L} \to  F_{ij}^{H,R}$ in the coefficient
of $f$, and $\rho^{F \dagger} \to \rho^{F}$
in the coefficient of $g$.

Finally, there is a two-loop contribution involving
a $W$-loop; the third diagram of figure \ref{fig:1loop}
is one of the many  that contribute. In the
CP-conserving  2HDM considered here, $A$  does not couple to the
$W$ or the goldstones, so cannot
appear in  these diagrams. CHK 
compute separately  the diagrams with 
either photon or $Z$ exchange between the lepton and $W$
(see figure  \ref{fig:1loop}). The $Z$-mediated amplitude
is proportional to $1 - 4 \sin^2 \theta_W$,
 and estimated by CHK to be about 10\% of
the photon-mediated amplitude,
so only the
the $\g$-amplitude is considered here.
The $H$ and $h$ contributions individually  are:
\bea
-2 \sqrt{2} G_F m_\mu A_L^{W~loop}& = &\pm \frac{e \alpha }{32 \pi^3}  
\frac{\Lambda_6 v}{M^2}
\frac{\rho^{E}_{ \mu e}}{\sqrt{2}}
\left( 3f(z_\phi) + 5 g(z_\phi) + \frac{3}{4}[g(z_\phi) +h(z_\phi)] + 
\frac{f(z_\phi) -g(z_\phi)}{2z_\phi }\right) 
\eea
where 
 $z_\phi = m_W^2/m_\phi^2$, and  - is for $\phi =H$, +is for $\phi =h$. 
The sum of the contributions, in the decoupling limit, 
is
\bea
-2 \sqrt{2} G_F m_\mu A_L^{W~loop}
&\simeq&
- \frac{e \alpha }{64 \pi^3}  \frac{\Lambda_6 v}{M^2}
\frac{\rho^{E}_{ \mu e}}{\sqrt{2}}
\left[
\left(   \frac{ m_W^2}{M^2} \ln \frac{ m_W^2}{M^2}
\left(
\frac{35}{4} \ln \frac{ m_W^2}{M^2}
 + \frac{3}{2} \right) + 
\ln \frac{ m_W^2}{M^2} +2
\right)\right.  \nonumber \\
&&\left.
-
2 \left( 3f(z_h) + 5 g(z_h) + \frac{3}{4}[g(z_h)+h(z_h)
+ \frac{f(z_h) -g(z_h)}{2z_h }\
 ] \right) \right]
\label{fonctionneWW}
\eea
where the limiting forms of  eqn (\ref{idfg}) were used.
 The  first line of the approximate equality is
the contribution of $H$, and the last line  is
the contribution of $h$ where $z_h = m_W^2/m_h^2$,
and the parenthese evaluates to $\sim 7$.
The heavy $H$ exchanged  between the
lepton line and a goldstone loop generates an ${\cal O}(\log/M^2)$
term, which is carefully discussed by
CHK,  because the $1/M^2$ suppression
  arises from the
decoupling limit of $\cbma$, given in eqn (\ref{cbmadl}).
CHK  take the mixing angle
$\cbma$ as a free parameter,
 so  refer to this term as a ``non-decoupling''
contribution.

\subsubsection{ The relative importance of the ${\cal O}(1/M^2)$
and ${\cal O}(1/M^4)$ terms}
\label{sssec:relative}

Since EFT is an expansion in operator
dimension, it is interesting to know under what
conditions the  ${\cal O}(1/M^2)$ terms give
a good approximation to the full answer. 
These terms should arise in a relatively simple  EFT using
dimension six operators; 
a more extended
EFT would be required to reproduce the
  ${\cal O}(1/M^4)$ terms.
So  this section compares the
magnitude  of the ${\cal O}(1/M^2)$
and ${\cal O}(1/M^4)$ terms in 
the four classes of diagrams (one-, $b$-, $t$-  and  $W$-loop),
with little attention to signs and 2s.

For 
 the one-loop
 diagrams,  eqns 
(\ref{1ldc},\ref{1lh}) 
give the 
ratio of the   ${\cal O}(1/M^4)$ 
parts    to the   ${\cal O}(1/M^2)$ part
as
\bea
\left(
\frac{   v^2}{ M^2  }
 \right)
\frac{  
2 [\rho^E]_{\mu \mu}     \Lambda^2_6 
+ \left[   Y^E_{\mu \mu}  \Lambda_6   
- [\rho^E]_{\mu \mu} \Lambda_5 
 \right]( 1 +   \ln \frac{m_h^2}{M^2} /\ln \frac{m_\mu^2}{m_h^2} )
}
{
2 Y^E_{\mu \mu} \Lambda_6 
}
\eea
where I used $v^2/m_h^2 \simeq 4$. 
Recall that in this paper, $\tan \beta$ is defined 
from the leptons, see eqn (\ref{tanb}),
so ${[\rho^E]_{\mu \mu}}/{[Y^E]_{\mu \mu}} \equiv \tan \beta$.
Therefore, the ${\cal O}(1/M^2)$ parts  are larger than the  ${\cal O}(1/M^4)$ 
contributions, provided that
\footnote{There is an  ${\cal O}(1/M^4)$  term from
the heavy Higgs exchange diagrams that benefits
from an additional log enhancement --- it runs from
$M \to m_\mu$ rather than from  $m_h \to m_\mu$ like the 
 ${\cal O}(1/M^2)$ light higgs contribution. However,
this is insignificant, because $M/m_h \ll m_h/m_\mu$.}  
\beq
\Lambda_6  \tan \beta  \frac{v^2}{M^2} < 1 
~~~,~~~
\tan \beta  \frac{\Lambda_5}{ 4 \Lambda_6} \frac{v^2}{M^2} < 1 
~~~\Rightarrow  {\cal O}(\frac{1}{M^2}) > {\cal O}(\frac{1}{M^4}) 
\label{1l2>4}
\eeq
If  $1/50 < \tan \beta < 50$,  $v^2/M^2 < 0.01$,
and $\Lambda_5 \sim \Lambda_6 <1$,
then the  ${\cal O}(1/M^2)$  terms give
the correct order of magnitude, but 
the  ${\cal O}(1/M^4)$ terms can be required
 to get  two significant
figures for large  $\tan \beta$.

For the top loop, the ratio of
  ${\cal O}(1/M^4)$   over  ${\cal O}(1/M^2)$ 
  terms is
\bea
\left(
\frac{ v^2}{M^2} \right)  
\frac{
    Y^{U}_{tt} \Lambda_6    
}{[\rho^{U}]_{tt}  - 
  Y^{U}_{tt} \Lambda_6/\ln^2 \frac{m_t^2}{M^2}  
  }
\label{trapport}
\eea
where   $f(\frac{m_t^2}{m_h^2}) \simeq 1$ was used.
This shows that the  ${\cal O}(1/M^2)$  terms are
always larger (neglecting the possible
cancellation of $[\rho^{U}]_{tt} \ln^2 \frac{m_t^2}{M^2}$ against 
 $ \Lambda_6$), however, in 2HDMs 
where the top couples mostly to
the SM Higgs such that $[\rho^{U}]_{tt} \ln^2 \frac{m_t^2}{M^2} \ll  \Lambda_6$,
the dimension eight contribution is
only suppressed by $ \frac{m_t^2}{M^2} \ln^2 \frac{m_t^2}{M^2}$
which is $\simeq .2$ for $\frac{m_t^2}{M^2}  = 0.01$.
So the dimension eight contribution
can be numerically relevant in some areas
of parameter space.

For the $b$ loop, the ratio of
  ${\cal O}(1/M^4)$   over  ${\cal O}(1/M^2)$ 
  terms is
\bea
\left(
\frac{ v^2}{M^2} 
\right)
\frac{ 
 \left[
4 \Lambda_6
\rho^{D \dagger}_{bb} +
\left( Y^{D}_{bb}  
- \rho^{D}_{bb}\frac{\Lambda_5 }{\Lambda_6}
\right) \left(1 + \ln^2 \frac{m_h^2}{M^2}/\ln^2 (\frac{m_b^2}{m_h^2}\right) 
 \right]  }{4 Y^{D}_{bb}  
  }
\label{brapport}
\eea
So  the ${\cal O}(1/M^2)$ parts  are larger than the  ${\cal O}(1/M^4)$ 
contributions, provided that
\beq
 \frac{\Lambda_5 }{4 \Lambda_6 } \frac{\rho^D_{bb}}{Y^D_{bb}}  \frac{v^2}{M^2} < 1
~~~,~~~
\Lambda_6  \frac{\rho^D_{bb}}{Y^D_{bb}}  \frac{v^2}{M^2} < 1
~~~\Rightarrow  {\cal O}(\frac{1}{M^2}) > {\cal O}(\frac{1}{M^4}) 
\label{bl2>4}
\eeq
where  $\frac{\rho^D_{bb}}{Y^D_{bb}} \sim 35$
in 2HDMs where $\rho^D_{bb} \simeq 1$.
As in the case of the one-loop contribution,
the  logarithm multiplying the 
$ {\cal O}({1}/{M^4}) $ heavy Higgs exchange diagrams   runs
 from
$M \to m_b$ rather than from  $m_h \to m_b$, which
could give a factor 2.

For the $W$ loop, the ratio of
  ${\cal O}(1/M^4)$   over  ${\cal O}(1/M^2)$ 
  terms is
\bea
\frac{   \frac{ m_W^2}{M^2} \ln \frac{ m_W^2}{M^2}
\left(
\frac{35}{4} \ln \frac{ m_W^2}{M^2}
 - \frac{3}{2} \right)}{ 
\ln \frac{ m_W^2}{M^2} -12}
\sim   \frac{ m_W^2}{M^2} \ln^2 \frac{ m_W^2}{M^2}
\label{rapportW}
\eea
where the estimate neglects cancellations, and  shows
that the   ${\cal O}(1/M^4)$  contribution is only mildly
suppressed, because $z \ln^2 z$ does not decrease
rapidly.

So in summary,
provided that
\beq
\Lambda_6 \sim \Lambda_5~~,~~~
\frac{v^2}{M^2} \cot \beta  <1~~~,~~~ \frac{v^2}{M^2} \tan \beta  <1
\eeq
 in all four  types
of 2HDM,
then
the ${\cal O}(1/M^2)$  terms give
the correct order of magnitude, but 
the  ${\cal O}(1/M^4)$ terms can be required
 to get  two significant
figures  in some areas of parameter space.

\subsubsection{ The relative size of the four classes
of diagram}
\label{sssec:sizes}
 The previous section
suggested that it would be reasonable to use
an EFT with dimension six operators. 
So  the next question is to determine  the loop order to which
the matching and running should be
performed in the EFT,  in order  to reproduce
 the numerically dominant  ${\cal O}(1/M^2)$  
terms. So below are compared 
the magnitudes of the
 ${\cal O}(1/M^2)$  terms of the four
classes of diagrams.

The  ${\cal O}(1/M^2)$  part of the one-loop
contribution arises from $h$ exchange:
\bea
2\sqrt{2} G_F m_\mu A^{h,1~loop}_{L}&   \sim &
\frac{e m_\mu}{ \pi M^2} 
[\rho^E]_{\mu e}
Y^E_{\mu \mu} \Lambda_6  
\label{1lhM2}
\eea
where $v^2/m_h^2 \simeq 4$,  and
$\frac{1}{16 \pi} \ln \frac{m_\mu^2}{m_h^2}  
=0.998 \simeq 1$.

Normalising the  ${\cal O}(1/M^2)$  part of the two-loop
diagrams to eqn (\ref{1lhM2}) gives
\bea
{t~loop}
 & \sim &\frac{ \alpha m^2_t }{32\pi^2 m^2_\mu}  \frac{ 3Q_t^2}{\Lambda_6}
\left[   [\rho^{U}]_{tt}  \ln^2 \frac{m_t^2}{M^2}  
 -    \Lambda_6 ) \right]
\label{tdlrap}
\\
{b~loop}
 &\sim &  \frac{\alpha m^2_b }{16\pi^2 m^2_\mu}    3Q_b^2
\ln^2 (\frac{m_b^2}{m_h^2})  
\label{bdlrap}
\\
{W~loop}& 
\sim&
 \frac{ \alpha }{64 \pi^2}  \frac{ m^2_t}{m^2_\mu}
\left[
\ln \frac{ m_W^2}{M^2} -12 \right]
\label{Wrap}
\eea
where  $f(m_t^2/m_h^2) \simeq 1$ was used. Since
$ \frac{ \alpha }{64 \pi^2}  \frac{ m^2_t}{m^2_\mu} \simeq 32$,
this reproduces the well-known dominance of
the top and $W$ loops.  Interestingly, only the top loop
can be enhanced (or suppressed)  by  the angle $ \beta$ ---
the relative magnitude of the other terms is
mostly controlled by Standard Model parameters.

The two-loop top and $b$ contributions  contain log$^2$ terms, 
which  should arise at second order  in the one-loop
RGEs of dimension six operators. However, there
 are significant terms of the top and $W$
contributions without a log,  which presumably 
arise in two-loop matching. And the log
term of the $W$ amplitude should be generated by
 two-loop RGEs.  So we see that a two-loop
analysis would be required to reproduce the
dominant ${\cal O}(1/M^2)$ terms.


\section{The EFT version}
\label{sec:EFT}

The aim of this section is to obtain, in EFT, the 
``leading'' ${\cal O}([\alpha \log]^n/M^2)$ parts
of the $\meg$ amplitude. Appendix \ref{ssec:dim8}  discusses where
some  other parts of the full-model calculation
would arise.  EFT is transparently
reviewed in \cite{Georgi},
and the EFT construction here attempts to  follow
the recipe given there.
For simplicity, the EFT  has only    three
scales:  the  heavy Higgs scale $M$,
the weak scale $m_W$ (taken $\simeq m_Z, m_h,m_t$)
and a low scale $m_\mu$ (taken 
$\simeq  m_b$). Between $M$ and $m_W$, the
theory and operators are $SU(3) \times SU(2)
\times U(1)$ invariant; below $m_W$, they
are $SU(3) 
\times U(1)$ invariant.

The EFT constructed here
contains  dimension six operators,
which should reproduce the  ${\cal O}(1/M^2)$ terms
of the $\meg$ amplitude. 
In eqns (\ref{dip})-(\ref{yu6L}) are listed the  dimension six,
SU(2)-invariant operators
required above $m_W$. Eqn (\ref{6ou8}) 
gives some additional
 SU(3)$\times$U(1)-invariant,
dimension six
 operators which
are required below $m_W$
(but which  would have been dimension eight
in the  SU(2)-invariant formulation appropriate above
$m_W$). As expected,  the operator basis below $m_W$
should include   all four-fermion operators
that are  of  dimension six  
in a  QED$\times$
QCD-invariant theory.
If instead, the basis  is
restricted to SU(2)-invariant ``Buchmuller-Wyler'' operators, 
one cannot obtain  all the ${\cal O}([\alpha \log]^n/M^2)$ terms
of the $\meg$ amplitude (only the top loop is included).

The  EFT  studied here is at
``lowest order'' in the loop expansion: tree-level matching
of operator coefficients, and one-loop
RGEs.  This  should
reproduce the   ${\cal O}(\alpha^n \log^n)$ terms
\footnote{ Since the dipole operator has  only
two fermion legs but an  external photon,
the anomalous dimension mixing four-fermion operators
into the dipole  is  $\propto 1/e$. So in
counting  powers of $\alpha \log$, sometimes
one should  multiply by $e$.} in the amplitude.
I do not calculate the two loop matching,
or the two-loop RGEs, which
would allow to reproduce the dimension
six contributions of the top and $W$.

\subsection{Setting up   the EFT calculation}

The aim of this top-down EFT calculation  is
to reproduce the amplitude for $\meg$ in the 2HDM.
So the first step is to match out the heavy Higgses
 $H$ and $A$ at   the ``New Physics scale'' $M$.  
The $H^\pm$ are neglected  because 
 the full-model calculation (to which the EFT is compared) includes
only neutral Higgses. Presumeably  the  $H^\pm$ contribution
ensures SU(2) invariance. 
 The 2HDM  from above
$M$ is matched onto 
the (unbroken) Standard Model, with its full particle content,
and a selection of $SU(3) \times SU(2) \times U(1)$
invariant   dimension six operators. A list of possible operators 
was given by Buchmuller and Wyler \cite{BW}, 
and slightly reduced in \cite{Polonais}. Here,
only those operators which are required to
reproduce  the ${\cal O}(1/M^2)$ part
of the $\meg$ amplitude are selected. In matching out, 
the operator coefficients of  the effective theory
are assigned so as to  reproduce the tree-level Greens 
functions of the full theory, at zero external momentum.

 The second step is to run the operator coefficients
down to the scale $m_W$.
This running 
 should be performed with electroweak RGEs,
which  are given in \cite{PS,JMT}.However,
since CHK separate photon and $Z$ diagrams, 
and only the  photon contributions were retained 
in the previous section,
the running  
from $M\to m_W$ is performed with 
the RGEs of QED.  The operator
coefficients evolve\cite{HisanoEDM} with scale $\mu$ as
\beq
\mu \frac{\partial}{\partial \mu} \vec{C} =  \frac{\alpha_{em}}{4\pi}\vec{C} \bm{\Gamma} 
\label{RGE}
\eeq 
where the  operator coefficients have  been organised into a row vector
$ \vec{C} $,
and 
 $\frac{\alpha_{em}}{4\pi} \bm{\Gamma}$ is
the anomalous dimension matrix.
  The algorithm to calculate
$ \bm{\Gamma}$ is given, for instance, in \cite{BurasHouches}.
For a square
$\bm{\Gamma}$, this equation can be
perturbatively solved to give the  components of the vector $\vec{C}(\mu)$ 
at a lower scale $\mu$:
\bea
 C_A(M)\left(
\delta_{AB} - \frac{\alpha (\mu) }{4\pi} \left[ \Gamma\right]_{AB}
\ln \frac{M}{\mu}
 + 
\frac{\alpha ^2(\mu) }{32\pi^2} \left[ \Gamma \Gamma\right]_{AB}
\ln ^2  \frac{M}{\mu} +..
\right) =  C_B (\mu)
\label{opmix1l}
\eea

 At $m_W$,  the $W,Z,h$ and $t$ should be matched out.
The theory below $m_W$ should be 
$QED$ and $QCD$  for all the SM fermions except the top,
augmented by a complete set
of $QED \times QCD$-invariant dimension six operators.
For simplicity, I  consider only QED for the $b,\mu$ and $e$,
plus those dimension-six operators required in
matching onto the  tree-level 
Greens functions  of the SU(2)-invariant EFT from above $m_W$.

Finally,  the operator coefficients
are run  down to $m_\mu$, where the dipole
coefficient can be used to calculate the $\meg$ amplitude.
In principle,  the  RGEs of QED and QCD should be used. However,
 QCD  is neglected because  it is not
included in the full-model calculation of CHK.

\subsection{Matching at $M$ 
and one-loop running to $m_W$}

The operator for $\meg$ is given in eqn (\ref{Lmeg}). It is convenient
to rescale the coefficient as 
\beq
 2\sqrt{2} G_F m_\mu A^{ij}_R  (\overline{e}_i  \sigma^{\a \b}P_R  e_j ) F_{\a \b}
 = \frac{ Y^E_{\mu \mu}}{M^2} C^{ij}_{D,R} 
(\overline{L}_i  H_1 \sigma^{\a \b}P_R  E_j ) F_{\a \b}
\equiv \frac{ Y^E_{\mu \mu}}{M^2} C^{ij}_{D,R}  
{ O}^{ij}_{D,R}
\label{dip}
\eeq
so that the dimensionful operator coefficient
is suppressed by the heavy Higgs scale $M$.
 Above $m_W$, there is a hypercharge dipole and
an SU(2) dipole; here only the linear combination
corresponding to   the photon dipole is used, because  
only the QED part of
the SU(2) running is included.
To obtain
the ${\cal O}([\alpha \log]^n/M^2)$ parts
of the $\meg$ amplitude, only  three  additional  operators
from the  pruned Buchmuller-Wyler list
are required between $M$ and $m_W$:
\bea
{ O}^{e \mu tt}_{LE Q U}&=& (\overline{L}_e^A  E_\mu )\epsilon_{AB} 
(\overline{Q}^B_t U_t )
~~~~~~~~~~~~~~~~~~~~~~~~
\nonumber \\
{ O}^{ \mu e tt}_{LE Q U}&=& (\overline{L}_\mu^A  E_e )\epsilon_{AB} 
(\overline{Q}^B_t U_t ) 
\label{scalarmue}\\
{ O}^{e \mu tt}_{T, LEQU} &=&  (\overline{L}_e^A \sigma^{\mu\nu} E_\mu )\epsilon_{AB} 
(\overline{Q}^B_t \sigma_{\mu\nu} U_t )
\nonumber\\ 
{ O}^{\mu e tt}_{T, LEQU} &=&  (\overline{L}_\mu^A \sigma^{\mu\nu} E_e )\epsilon_{AB} 
(\overline{Q}^B_t \sigma_{\mu\nu} U_t )
\label{tensormue}  \\
{ O}^{e \mu}_{eH } &=&  H_1^\dagger H_1 \overline{L}_e H_1  E_\mu 
\nonumber\\ 
{ O}^{\mu e}_{e H } &=&  H_1^\dagger H_1 \overline{L}_\mu H_1  E_e 
 \label{yu6L} 
\eea
where  $\sigma $ is the anti-symmetric tensor $\frac{i}{2}[\g,\g]$,
and $\epsilon$ provides an SU(2) contraction.  
These operators appear in the Lagrangian as
\beq
\delta {\cal L} = - \sum_{operators}\frac{C^{ijmn}_{Z}}{M^2} { O}^{ijmn}_{Z} +h.c. 
\eeq
so the coefficients $C$ are dimensionless, and
the four fermion operators are normalised such that the Feynman rule is
$-iC/M^2$.

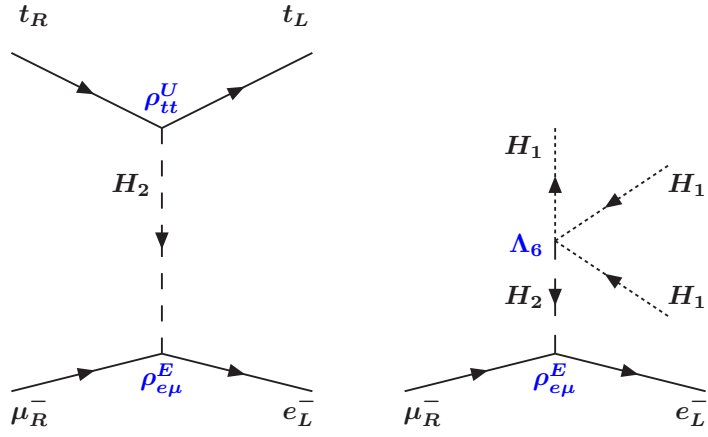
\begin{figure}[h]
\unitlength.5mm
\SetScale{1.418}
\begin{boldmath}
\begin{center}
%
%
\begin{picture}(60,100)(0,0)
\ArrowLine(0,0)(40,10)
\ArrowLine(40,10)(80,0)
\ArrowLine(0,90)(40,70)
\ArrowLine(40,70)(80,90)
\DashArrowLine(40,70)(40,10){4}
\Text(10,-5)[r]{$\mu_R^-$}
\Text(72,-5)[l]{$e_L^-$}
\Text(10,100)[r]{$t_R$}
\Text(72,100)[l]{$t_L$}
\Text(37,55)[r]{$H_2$}
\Text(40,78)[c]{${\color{blue}\rho^U_{tt}}$}
\Text(40,3)[c]{${\color{blue}\rho^E_{e \mu  }}$}
\end{picture}
\qquad \qquad \qquad
\begin{picture}(60,100)(0,0)
\ArrowLine(0,0)(40,10)
\ArrowLine(40,10)(80,0)
\DashArrowLine(40,40)(40,70){1}
\DashArrowLine(40,40)(40,10){4}
\DashArrowLine(70,60)(40,40){1}
\DashArrowLine(70,20)(40,40){1}
\Text(10,-5)[r]{$\mu_R^-$}
\Text(72,-5)[l]{$e_L^-$}
\Text(70,55)[l]{$H_1$}
\Text(70,25)[l]{$H_1$}
\Text(37,25)[r]{$H_2$}
\Text(37,65)[r]{$H_1$}
\Text(40,3)[c]{${\color{blue}\rho^E_{e \mu }}$}
\Text(37,39)[r]{${\color{blue}\Lambda_6}$}
\end{picture}
\end{center}
\end{boldmath}
\vspace{10mm}
\caption{The left 
diagram generates the  dimension six $(\overline{Q}_3  U_t) 
(\overline{L}_e  E_\mu)$ operator, by
 matching-out  the  heavy doublet Higgs $H_2$ (dashed line).
The right diagram generates the dimension six
$H_1^\dagger H_1 (\overline{L}_e H_1  E_\mu)$ 
operator.
\label{fig:2} }
\end{figure}

The last pair of  operators, eqn (\ref{yu6L})  will give the flavour-changing
light-Higgs interaction required for the   one-loop, $b$-loop
and $W$-loop amplitudes. As can be seen from the  right diagram
 of figure \ref{fig:2}, matching  at 
the scale $M$ of  the tree-level Greens functions
  of the full theory onto  those of the SM + dimension six operators
gives coefficients
\beq
 \frac{C_{eH}^{e\mu}}{M^2} = -\frac{[\rho^E]_{e \mu} \Lambda_6}{M^2}
~~~,~~~
 \frac{C_{eH}^{\mu e}}{M^2} = -\frac{[\rho^E]_{ \mu e} \Lambda_6}{M^2} ~~~.
\eeq
The running of these coefficients
between $M$ and $m_W$ is neglected,   because it is not required to
reproduce the CHK results.

 The first four operators  of the list above
will generate the top loop contribution.
Matching at $M$ via the diagram given on
the left in figure \ref{fig:2} gives coefficients for the scalar $LEQU$
operators
\beq
\frac{C_{LEQU}^{e\mu tt}}{M^2} = -\frac{[\rho^E]_{e \mu} [\rho^{U \dagger}]_{tt}}{M^2}
~~~,~~~
\frac{C_{LEQU}^{\mu e tt}}{M^2} = -\frac{[\rho^E]^*_{ \mu e} [\rho^{U \dagger}]_{tt}}{M^2} ~~~,
\eeq
where the negative sign is from the scalar propagator.
Then, between  the scales $M$ and $m_t$,
the RGEs of QED mix the  scalar operator 
${ O}^{e \mu tt}_{LE Q U}$
 to the tensor 
${ O}^{e \mu tt}_{T, LEQU}$,
 and the tensor to the dipole,
such that the coefficient of ${ O}^{e\mu}_{D,L}$ is
\beq
-m_\mu   \frac{C_{D,L}^{e\mu}}{M^2}  = 
 -\frac{C^{e \mu tt}_{LE Q U}}{M^2} \frac{e \alpha  }{128\pi^3} \left[
(2Q_t)(8N_c Q_t m_t)
\right]
\log ^2  \frac{M}{m_t}
=  \frac{e \alpha }{32\pi^3M^2} 
 3 Q^2_t m_t 
[\rho^E]_{e \mu} [\rho^{U \dagger}]_{tt}
\log ^2  \frac{m_t^2}{M^2}
\label{topD}
\eeq
where in the brackets after the first equality
is the product of the scalar$\to$tensor
 and tensor$\to$dipole elements of the
anomalous dimension matrix $\bm{\Gamma}$. This  agrees
with the ${\cal O}(1/M^2)$ part of   eqn (\ref{fonctionnett})
that is generated by heavy Higgs exchange.

\subsection{Matching 
at $m_W$ and 
one-loop running to $m_\mu$}

At the weak scale $ \simeq m_W$, the
$h, W,$ and $t$ should be matched out of the theory,
onto a basis of dimension six operators. These
operators  should 
respect the  gauge symmetries
below $m_W$, which,  in the absence of
the $h,W,$ and $Z$,   are
 QCD and QED. So there is no reason to impose
SU(2) on the operator basis below $m_W$. 

\ben
\item 
If the matching is performed at tree or one loop, 
then  matching out the top leaves only  the contribution
to the dipole operator given in eqn (\ref{topD}). 
The top-loop contribution with a light higgs,
which is ${\cal O}(1/M^2)$,
could  be included   by matching at two-loop.

\item
In matching out the Higgs $h$,
the one-loop, and $b$-loop   ${\cal O}(1/M^2)$-contributions
can be obtained  by
matching onto the  scalar and tensor operators:
\bea
{ O}_{S}^{e\mu bb} = (\overline{e} P_R \mu)   (\overline{b}  P_Rb)
& & 
{ O}_{S}^{\mu  e bb} =
(\overline{\mu} P_R e) 
(\overline{b} P_R b) 
\nonumber \\
{ O}_{T}^{e\mu bb} = (\overline{e} \sigma P_R \mu) 
(\overline{b } \sigma  P_R b) 
&&
 { O}_{T}^{\mu  e bb} = (\overline{\mu} \sigma P_R e) 
(\overline{b } \sigma  P_R b) 
\nonumber \\
{ O}_{S}^{e \mu    \mu \mu} =
(\overline{e} P_R\mu) 
(\overline{\mu} P_R \mu) 
&&
{ O}_{S}^{\mu  e  \mu \mu} = 
(\overline{\mu} P_R e) 
(\overline{\mu} P_R \mu)
\nonumber \\
{ O}_{T}^{e \mu    \mu \mu} 
= (\overline{e} \sigma P_R\mu) 
(\overline{\mu} \sigma  P_R\mu) 
&&
{ O}_{T}^{\mu  e  \mu \mu} = (\overline{\mu} \sigma P_R e) 
(\overline{\mu} \sigma  P_R \mu) 
\label{6ou8}
\eea
If  $SU(2)$ were imposed, these operators
would be  of dimension eight (for instance,
the first operator could be written as
$^{(8)}{ O}^{e \mu  bb }_{LEQD} =  (\overline{L}_e H E_\mu) 
(\overline{Q}_3 H  D_b)$). However, they
are  of dimension six in the QCD$\times$QED-invariant
EFT below $m_W$, and 
are required in the 2HDM to correctly reproduce
the ${\cal O}(\alpha^n \log^n/M^2)$ terms that
dimension-six, one-loop EFT should obtain.
 The operators of eqn (\ref{6ou8})
are  not included  in  the EFT  analysis of $\meg$ performed by
Pruna and Signer \cite{PS},  who restrict to  dimension six
SU(2)-invariant operators.

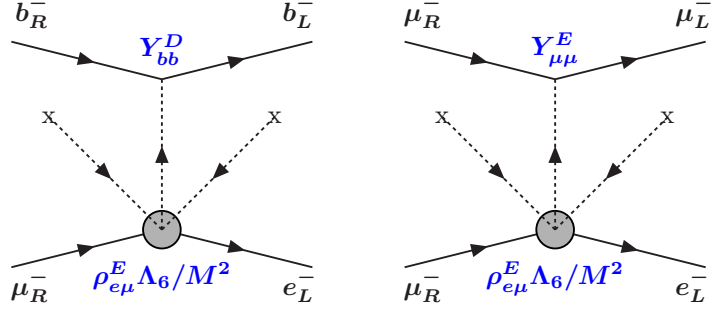
\begin{figure}[h]
\unitlength.5mm
\SetScale{1.418}
\begin{boldmath}
\begin{center}
\begin{picture}(60,100)(0,0)
\ArrowLine(0,0)(40,10)
\ArrowLine(40,10)(80,0)
\ArrowLine(0,60)(40,50)
\ArrowLine(40,50)(80,60)
\GCirc(40,10){5}{.7}
\DashArrowLine(40,10)(40,50){1}
\DashArrowLine(70,40)(40,10){1}
\Text(70,40)[c]{x}
\Text(10,40)[c]{x}
\DashArrowLine(10,40)(40,10){1}
\Text(10,-5)[r]{$\mu_R^-$}
\Text(72,-5)[l]{$e_L^-$}
\Text(10,68)[r]{$b_R^-$}
\Text(72,68)[l]{$b_L^-$}
\Text(40,58)[c]{${\color{blue}Y^D_{b b}}$}
\Text(40,-3)[c]{${\color{blue}\rho^E_{e \mu } \Lambda_6/M^2}$}
\end{picture}
\qquad \qquad \qquad
\begin{picture}(60,100)(0,0)
\ArrowLine(0,0)(40,10)
\ArrowLine(40,10)(80,0)
\ArrowLine(0,60)(40,50)
\ArrowLine(40,50)(80,60)
\GCirc(40,10){5}{.7}
\DashArrowLine(40,10)(40,50){1}
\DashArrowLine(70,40)(40,10){1}
\DashArrowLine(10,40)(40,10){1}
\Text(70,40)[c]{x}
\Text(10,40)[c]{x}
\Text(10,-5)[r]{$\mu_R^-$}
\Text(72,-5)[l]{$e_L^-$}
\Text(10,68)[r]{$\mu_R^-$}
\Text(72,68)[l]{$\mu_L^-$}
\Text(40,58)[c]{${\color{blue}Y^E_{\mu \mu}}$}
\Text(40,-3)[c]{${\color{blue}\rho^E_{e \mu } \Lambda_6 /M^2}$}
\end{picture}
\end{center}
\end{boldmath}
\vspace{10mm}
\caption{The right 
  diagram generates  the   QCD$\times$QED invariant, dimension six  operator
$(\overline{e } P_R \mu)(\overline{\mu} P_R \mu)$, by
 matching-out  the  light  Higgs $ h$ (dotted line),
which has a SU(2)-invariant dimension-six LFV  vertex represented
by the grey circle. The free Higgs legs attach to
the vev.
The left diagram generates a similiar  operator
$(\overline{b} P_R b) 
(\overline{e} P_R \mu)$  involving $b$ quarks.
\label{fig:dim8} }
\end{figure}

From the diagrams illustrated in figure \ref{fig:dim8},
one obtains
\bea
 \frac{C^{e \mu  \mu  \mu   }_{S}}{M^2}  =  \frac{[Y^E]_{\mu \mu}[\rho^E]_{ e\mu } \Lambda_6v^2/2}{ m_h^2 M^2} &&
 \frac{C^{ \mu e  \mu  \mu }_{S}}{M^2} =  \frac{[Y^E]_{\mu \mu}[\rho^E]_{ \mu e } \Lambda_6v^2/2}{ m_h^2 M^2} \\
 \frac{C^{e \mu  bb }_{S}}{M^2} =  \frac{[Y^D]_{bb}[\rho^E]_{ e\mu } \Lambda_6v^2/2}{ m_h^2 M^2} &&
 \frac{C^{ \mu e  bb }_{S} }{M^2}=  \frac{[Y^D]_{bb}[\rho^E]_{ \mu e } \Lambda_6v^2/2}{ m_h^2 M^2} 
\eea
so the coefficients are ${\cal O}(1/M^2)$
because $v^2/m_h^2 \simeq 4$.

Then, between  the scales $m_W$ and $m_\mu$,
the RGEs of QED mix the   scalar operator 
${ O}^{e \mu bb}_{S}$
 to the tensor
${ O}^{e \mu bb}_{T}$, which mixes to the
dipole, exactly as in  the previously
discussed  case of tops. So  with the
anomalous dimensions as in eqn (\ref{topD}), the dipole  coefficient
\beq
-m_\mu  \frac{ C_{D,L}^{e\mu}   }{M^2}
=-\frac{C^{e \mu bb}_{S}}{M^2} \frac{e \alpha m_b  }{32\pi^3} 
 3 Q^2_b 
\log ^2  \frac{m_\mu^2}{m_h^2}
= -
\frac{e \alpha m_b  }{64\pi^3M^2} 
 3 Q^2_b   \frac{[Y^D]_{bb}[\rho^E]_{ e\mu } \Lambda_6v^2}{  m_h^2 }
\log ^2  \frac{m_\mu^2}{m_h^2}
\label{bD}
\eeq
is generated, in agreement with  the ${\cal O}(1/M^2)$ part
of the $b$-loop contribution to $\meg$, given in eqn (\ref{fonctionnebb}).

In matching out the Higgs $h$, the flavour-changing
operator ${ O}^{e \mu}_{eH}$ can also generate
the dimension six operator ${ O}^{ e \mu bb}_{LE  D Q}= 
(\overline{L}_e  E_\mu )  (\overline{D}_b Q_b )$ 
with coefficient $\propto \rho^E_{\mu e} \Lambda_6 v^2/(m_h^2 M^2)$.
However, this operator is not useful for generating
$\meg$  via one-loop RGEs, because there is
no tensor operator for it to mix to, on the  way to the dipole.
The reason there is no tensor, is that $\sigma$
and $\sigma \gamma_5$ are related: $\sigma_{\mu \nu} = \frac{i}{2} 
  \varepsilon_{\mu \nu \a \b} \sigma^{\a \b} \g_5$, which implies
that  
$(\overline{e}\sigma^{\a \b} \g_5 \mu) (\overline{b}\sigma_{\a \b} \g_5 b) 
=
 (\overline{e} \sigma_{\mu \nu} 
\mu) (\overline{b}\sigma^{\mu \nu}  b ) 
$ or $(\overline{e}\sigma^{\a \b} P_L \mu) 
(\overline{b}\sigma_{\a \b} P_R b)
= 0$.

The scalar operator with three muons, 
${ O}^{e \mu \mu \mu}_{S}$, 
mixes directly  to the dipole via
a penguin diagram, 
so the coefficient of ${ O}^{e\mu}_{D,L}$ is
\beq
-m_\mu \frac{ C_{D,L}^{e\mu} }{M^2}  = 
C^{ \mu e \mu \mu}_{S} \frac{\alpha }{16\pi^2} \left[
\frac{-1}{e}
\right]
\log   \frac{m_h}{m_\mu}
=   
 \frac{e m_\mu  }{64\pi^2 M^2} 
\frac{[Y^E]_{\mu \mu}[\rho^E]_{ \mu e } \Lambda_6v^2}{ m_h^2}
\log  \frac{m^2_\mu}{m_h^2}
\label{mmme}
\eeq 
where in the brackets is the  scalar$\to$dipole element of the
anomalous dimension matrix $\bm{\Gamma}$. This  agrees
with the ${\cal O}(1/M^2)$ part of the one-loop,
light higgs
diagrams   eqn (\ref{1lh}).

\item
In matching out the $W$ at tree or one-loop,  none of
the $W$-loop contribution to $\meg$ is included, because  the light
higgs part arises in 2-loop matching.
In the full-model
result of eqn (\ref{fonctionneWW}),
there is also  an  ${\cal O}(1/M^2)$ heavy higgs part,
but it  is only enhanced by one log,
and presumeably arises in the 2-loop RGEs.

\een


\section{ Discussion}
\label{sec:disc}

The  CP-conserving 2HDM  studied here is  a  minimal extension of the
Standard Model ---
it  has only
one LFV coupling $[\rho^E]_{\mu e} \simeq [\rho^E]_{ e \mu}$
(see eqn \ref{modeliiihqq}), 
the magnitude of the other new   flavoured couplings
is controlled by the  single parameter $ \tan \beta$,
defined in eqn (\ref{tanb}),
and the  only  new  mass scale is the heavy doublet mass $M$,
taken $ \gsim 10v$.
The one and two-loop contributions to
 $\meg$ of the neutral Higgses were calculated by
Chang, Hou and Keung\cite{CHK}, and their result,
in the decoupling limit,
is given in eqns  (\ref{1lh}),  (\ref{1ldc}),  
(\ref{fonctionnett}),  (\ref{fonctionnebb})  
and  (\ref{fonctionneWW}).

Section \ref{sec:EFT} tries  to reproduce  the amplitude for $\meg$, in a
 simple EFT with three scales: $M, m_W, m_\mu$. At the scale $M$, the
heavy Higgses were ``matched out'' onto dimension six, SU(2)-invariant
operators. Between $M$ and $m_W $, the
operator coefficients should run according to
electroweak RGEs, but I used those of QED (because I
compare only to the photon diagrams of \cite{CHK}). At $m_W$, the
$W,Z,h$ and $t$ are matched out, so below $m_W$ is an EFT
containing all Standard Model fermions but the top, interacting 
via  QCD, QED and various four-fermion  operators. Finally
the operator coefficients  run  to
$m_\mu$ according to the RGEs of QED.

 There were three issues in
reproducing  the amplitude for $\meg$ in EFT:
\ben
\item In order to  obtain
the  ${\cal O}(1/M^2)$
terms of the full-model calculation, 
the operator basis below the weak scale needed to include 
  QED$\times$QCD-invariant
 four-fermion operators  which are dimension six, but
would have been dimension  eight in the
SU(2)-invariant  basis appropriate above $m_W$\footnote{It is reasonable to use a 
basis of   QED$\times$QCD-invariant
four-fermion operators below $m_W$, because 
the  EFT recipe \cite{Georgi} says that   one
should use a complete basis of   operators
consistent with the symmetries of the theory---
which, below $m_W$,  is 
 QED$\times$QCD  (but not SU(2)).}. 

This was the case for  the one-loop  contribution
of the  light higgs to $\meg$:  the light
higgs has a flavour-changing dimension six interaction
$H_1^\dagger H_1 \overline{L}_e H_1 E_\mu$.
At $m_h$, the  light higgs can be
matched onto  the  operator $v^2
(\overline{e} P_R \mu) 
(\overline{\mu} P_R \mu)$ = 
$(\overline{L}_e H E_\mu) 
(\overline{L}_\mu H  E_\mu)$ with coefficient $1/(M^2 m_h^2)$,
as illustrated on the right in figure \ref{fig:dim8}.
Then,  in QED
running  down to $m_\mu$,
this operator mixes to the dipole 
 by a penguin diagram obtained by closing a muon loop, 
inserting  a mass  and attaching a photon.
 So
the contribution to $\meg$ is $\propto v^2/(M^2 m_h^2) \log (m_\mu^2/v^2) 
\sim \log (m_\mu^2/v^2) /M^2 $.

However, although the  coefficients of 
$(\overline{e} P_R \mu) 
(\overline{\mu} P_R \mu)$
and
$(\overline{e} P_R \mu) 
(\overline{b} P_R b)$
are  formally ${\cal O}(1/M^2)$,  they are also   small,
because proportional to light fermion Yukawa couplings.
As discussed in section \ref{sssec:sizes}, 
they do not give a numerically relevant contribution
to $\meg$ in the decoupling limit of the  2HDM.
So  although these extra operators are required below $m_W$  to
obtain all the ${\cal O}(1/M^2)$ terms, 
they are not neccessary for getting a  reasonable
approximation to the answer\footnote{ This 
distinction  is maybe related to a study in $B$ physics \cite{AGC},
which gave the restrictions 
on the coefficients  of
 QED $\times$ QCD operators below $m_W$,  
 obtained by assuming that
New Physics can be described above the
weak scale  by  SU(2)-invariant, dimension six operators. 
The authors  conclude that  operators of the
structure   $(\overline{Q} H D) 
(\overline{Q} H  D)$  can be neglected ---
perhaps because the operator $H_1^\dagger H_1 \overline{Q} H_1 D$,
 was not included above $m_W$
(this operator is included in \cite{ACFG}),
or perhaps because   the Yukawa suppression makes
these operators irrelevant}.

\item    There are two-loop contributions
involving a top or $W$ loop,
initially discussed  by Bjorken and
Weinberg\cite{BW}, and 
illustrated
in figure  \ref{fig:1loop}.  As discussed in
section
\ref{sssec:sizes}, these  always 
dominate the one-loop contribution to $\meg$ in the decoupling
limit.   
 This disorder in the loop expansion
occurs because the one-loop contribution 
 is suppressed by the square of the  muon yukawa coupling,
whereas the two-loop diagrams are proportional to the
square of the gauge coupling or of the  top yukawa.
 In the EFT, 
 two-loop matching and running  would be  required to
reproduce the $W$-loop, and part of the top
loop.

This   ``disorder''
 is partly a feature of the dipole operator,
and partly of the 2HDM. 
The  dipole operator has a Higgs leg, and at one loop,
that Higgs can attach to the  fermion line, or
 to the boson line, if the
boson of the loop has a dimension three
coupling to the Higgs (such as the
$\mu^* H^*_u L E$ interaction in supersymmetry).
However, at two loop,  there are many
more possibilities for attaching the
external Higgs leg with an ${\cal O}(1)$
coupling.  In some models, it could be possible to
have ${\cal O}(1)$
coupling of the external Higgs leg to the
fermion or boson of the one-loop diagram.
However, in the case of the 2HDM,  there are no new fermions
so the fermion of the one-loop diagram is at best a $\tau$.
The boson is a Higgs without dimension three
interactions, so the one-loop diagram
is  suppressed by small Yukawa couplings.
This issue could be addressed by running
and matching at two-loop\footnote{The canonical EFT
recipe says that one should match at loop order $j$, 
and compute RGEs to loop order $j+1$, so as
to reproduce all terms up to order $\alpha^n \log^{n-j}$.
However, the dipole operator can  only be  generated
with loops, so ``leading order'' matching, for the
dipole, is  at one-loop rather than at tree.  
It is therefore debatable\cite{CNR} whether $j$-loop RGEs
should be combined with matching at order $j$
or $j-1$ for the dipole. In order to
reproduce the numerically significant parts of the
$\meg$ amplitude in the 2HDM, it seems one should
run and match at the {\it same} order, that is,
two-loop.}, as is done for $b\to s \g$ \cite{BurasHouches}.

\item  Section \ref{sssec:relative} checked that the
${\cal O}(1/M^2)$ terms in the amplitude 
are larger than the ${\cal O}(1/M^4)$ terms,
so a parametrisation in terms of dimension
six operators should work.  However, 
dimension eight operators can be enhanced by
logs and factors of $\tan \beta$, so that it may
be neccessary to include them for numerical accuracy.

It is convenient to neglect dimension eight operators, and they are
expected to be suppressed by $ \sim z =v^2/M^2$. However,
if dimension eight operators are log-enhanced
in running, and dimension six only
contribute in matching,  then dimension eight 
are only suppressed by $z \log^n(z)$. 
This was the case  in the 2HDM for the two loop contributions
of the $W$s  and $b$s. Since the
high scale is not so high in the 2HDM, ({\it e.g.} $v/M \simeq 1/10$),
$z =v^2/M^2 \simeq 0.01$, but $z \log^2(z) \simeq .2$.
 
There is   also the inevitable ignorance, in EFT, about 
the magnitude  of operator coefficients in
the full theory.  This uncertainty  is parametrised by
$\tan \beta$ in the 2HDM.  To justify neglecting
the dimension eight operators, the  restrictions
$\cot \beta, \tan \beta < v^2/M^2$ had to be imposed.
However, since  flavour physics is about 
the hierarchy of couplings, it may
not be sensible to assume that all New Physics couplings are ${\cal O}(1)$
at the New Physics  scale.

\een


\begin{figure}[ht]
\begin{center}

\begin{picture}(70,70)
\ArrowLine(0,60)(20,30)
\ArrowLine(20,30)(40,60)
\put(-6,60){$\mu$} 
\put(45,60){$e$} 
\GCirc(20,25){5}{.7}
\ArrowLine(0,-10)(20,20) 
\ArrowLine(20,20)(40,-10)
\put(-16,-12){$q$} 
\put(46,-10){$q$} 
\Photon(10,45)(10,5){2}{4}
\put(50,20){$+...~~~\rightarrow$}  
\end{picture}
\hspace{1.5cm}
\begin{picture}(70,70)
\ArrowLine(0,60)(20,30)
\ArrowLine(20,30)(40,60)
\put(-6,60){$\mu$} 
\put(45,60){$e$} 
\GCirc(20,25){5}{.7}
\put(17,10){$\sigma$} 
\put(17,35){$\sigma$} 
\ArrowLine(0,-10)(20,20) 
\ArrowLine(20,20)(40,-10)
\put(-16,-12){$q$} 
\put(46,-10){$q$} 
\put(50,20){$~~~\rightarrow$}  

\end{picture}
\hspace{1.5cm}
\begin{picture}(70,70)
\GCirc(20,37){5}{.7}
\ArrowLine(0,60)(20,40)
\ArrowLine(20,40)(40,60)
\put(18,46){$\sigma$} 
\put(18,26){$\sigma$} 
\put(-6,60){$\mu$} 
\put(45,60){$e$} 
\ArrowArc(20,20)(15,0,360)
\put(40,25){$q$} 
\Photon(20,5)(20,-10){2}{3} 
\end{picture}
\end{center}
\caption{ Obtaining the Barr-Zee diagrams at second order in
the one-loop RGEs: the scalar operator (grey circle of the left
diagram) is mixed to the tensor
(grey circle of the next diagram) via  photon  exchange between
the leptons and quarks (only one  the four diagrams is drawn). Then the quark
loop of the tensor operator is closed to mix it to the dipole.
 \label{fig:EFTbz}
}
\end{figure}
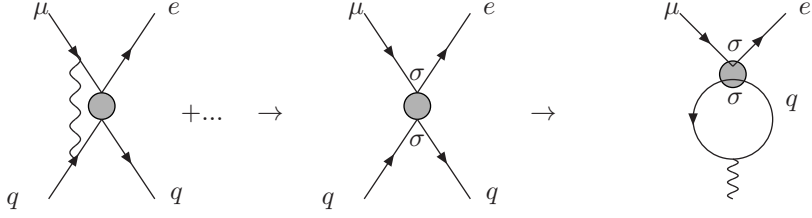

This exercise  located the ``Barr-Zee'' diagrams (see the
right two diagrams of figure \ref{fig:1loop}) in an EFT
description. 
\ben
\item The top loop with heavy Higgses contributes
at second order in the one-loop RGEs for dimension six
operators, as sketched in figure \ref{fig:EFTbz},
so is included in
the EFT used here. (There is also a dimension eight
contribution). The top loop with the light higgs
would appear in two-loop matching at $m_W$.
\item 
The $W$-loop with heavy Higgs contributes in
two-loop RGEs of dimension six operators, and
at second order in  the one-loop RGEs for dimension
eight operators.  The $W$-loop with the
light higgs would contribute in two-loop
matching at $m_W$. As a result, none of the $W$
contribution is included in the EFT used here.
\item  The $b$-loop contribution  is small
in the decoupling limit. 
The light higgs contribution, which appears
at  second order in the one-loop
RGEs for dimension six operators below $m_W$,
is suppressed by the   square
of the SM  Yukawa coupling of the $b$.
The heavy Higgs contribution, which could
be $\tan\beta$ enhanced,  however
only contributes at dimension eight,
via second order terms in the one-loop
RGEs. This is because the tensor
operator $(\overline{\psi_1} \sigma P_L \psi_2)
(\overline{\psi_3} \sigma P_R \psi_4)$  vanishes,
so the SU(2)-invariant dimension six scalar operator
that can be constructed with leptons and down quarks,
$   (\overline{Q}  D)(\overline{E}  L)$,
cannot mix via RGEs to a tensor then the dipole.
\een


\section{Summary}
This paper used Effective Field Theory (EFT) to calculate the amplitude for
$\meg$ in the decoupling limit of a 2 Higgs Doublet Model(2HDM) of Type III.
A ``leading order'' EFT was used, with one-loop running, tree matching,
dimension-six operators, and three scales (the heavy doublet mass $M$,
the electroweak scale $m_W$, and $m_\mu$). Comparing 
this EFT to  a calculation performed  in the 2HDM 
showed two things: first, that this EFT
reproduces, (as it should), the ${\cal O}([\alpha \log]^n/M^2)$ terms of the
full-model calculation, {\it provided that} the operator basis
below $m_W$ is constructed with dimension six $SU(3)\times U(1)$
invariant operators, (some of
 which could be dimension-eight in
an SU(2)-invariant construction).
Secondly, to obtain the numerically dominant contributions
to $\meg$ in the 2HDM, two-loop RGEs and two-loop matching are
required, but the enlarged operator basis below
 the weak scale is not.


The 2HDM also illustrates  that
higher dimensional  operators may  be numerically relevant,
because they can be enhanced by
unknown large couplings of the high-scale model ($e.g. \tan \beta$),
or by logarithms. 


\section*{Acknowledgements}

I am very grateful to Junji Hisano  for a conference invitation, 
and for relevant questions and extensive discussions
which motivated this work.   I thank Aielet Efrati, Marco Pruna and
Adrian Signer  for
careful reading and comments.

\appendix

\section{Translating between  notations}

To present the results of CHK, a translation dictionary between
their notation and the notation here  is useful.
CHK give the Lagrangian as
\beq
{\cal L} = -\overline{t}  
\frac{ m_t   }{v}
\left[ \Delta ^{\phi *}_{tt} P_R + \Delta ^{\phi}_{tt} P_L\right]t \phi
-\overline{e}  
\frac{\sqrt{m_e m_\mu  }  }{v}
\left[ \Delta ^{\phi R}_{e \mu} P_R + \Delta ^{\phi L}_{e \mu } P_L\right] \mu
+ gm_W \cos\theta_\phi W^+ W^- \phi
\eeq
where recall  $\frac{ m_t   }{v} =
\frac{ Y_{tt}^U   }{\sqrt{2}}$.
Defining  $\Delta ^{\phi *}_{tt} = \Delta ^{\phi R}_{tt}$,
 $\Delta ^{\phi }_{tt} = \Delta ^{\phi L}_{tt}$,  gives
\bea
\Delta ^{\phi X}_{ij} = 
 \frac{v}{\sqrt{ m_i m_j}} F^{\phi X}_{ij}
&& \cos\theta_\phi  = \left\{ 
\begin{array}{cc}
\cbma & \phi = H\\
\sbma & \phi = h\\
0 & \phi = A
\end{array}
\right.
\eea
CHK give their two-loop amplitudes $A_{L,R}$ in a different
normalisation from  eqn (\ref{BRmeg}); the relation is:
\beq
- A_{L,R}^{KO}= \frac{e \alpha}{64 \pi^3 } \sqrt \frac{m_e}{m_\mu}(A_{L,R}^{CHK})^*
\eeq
where the negative sign is because Kuno-Okada subtract 
their dipole operator from the SM Lagrangian, and
the hermitian conjugate is because Kuno-Okada
write an operator that mediates $\mu^+ \to e^+ \g$,
and CHK compute amplitudes for  $\mu^- \to e^- \g$.

\section{Finding the remaining terms in an  EFT calculation}
\label{ssec:dim8}

This appendix discusses where to find, in an EFT, the  missing parts
eqns  (\ref{1lh}),  (\ref{1ldc}),  (\ref{fonctionnett}),  (\ref{fonctionnebb})  
and  (\ref{fonctionneWW}).

\subsubsection{The full-model one-loop contribution}

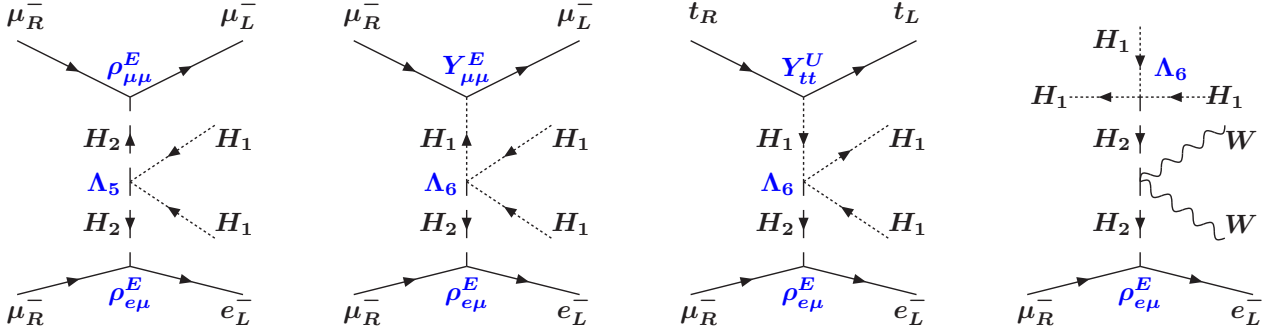
\begin{figure}[h]
\unitlength.375mm
\SetScale{1.064}
\begin{boldmath}
\begin{center}
%
%
\begin{picture}(60,100)(0,0)
\ArrowLine(0,0)(40,10)
\ArrowLine(40,10)(80,0)
\ArrowLine(0,90)(40,70)
\ArrowLine(40,70)(80,90)
\DashArrowLine(40,40)(40,70){4}
\DashArrowLine(40,40)(40,10){4}
\DashArrowLine(70,60)(40,40){1}
\DashArrowLine(70,20)(40,40){1}
\Text(10,-5)[r]{$\mu_R^-$}
\Text(72,-5)[l]{$e_L^-$}
\Text(10,100)[r]{$\mu_R^-$}
\Text(72,100)[l]{$\mu_L^-$}
\Text(70,55)[l]{$H_1$}
\Text(70,25)[l]{$H_1$}
\Text(37,25)[r]{$H_2$}
\Text(37,55)[r]{$H_2$}
\Text(40,80)[c]{${\color{blue}\rho^E_{\mu \mu}}$}
\Text(40,0)[c]{${\color{blue}\rho^E_{ e \mu}}$}
\Text(37,39)[r]{${\color{blue}\Lambda_5}$}
\end{picture}
\qquad \qquad \qquad
\begin{picture}(60,100)(0,0)
\ArrowLine(0,0)(40,10)
\ArrowLine(40,10)(80,0)
\ArrowLine(0,90)(40,70)
\ArrowLine(40,70)(80,90)
\DashArrowLine(40,40)(40,70){1}
\DashArrowLine(40,40)(40,10){4}
\DashArrowLine(70,60)(40,40){1}
\DashArrowLine(70,20)(40,40){1}
\Text(10,-5)[r]{$\mu_R^-$}
\Text(72,-5)[l]{$e_L^-$}
\Text(10,100)[r]{$\mu_R^-$}
\Text(72,100)[l]{$\mu_L^-$}
\Text(70,55)[l]{$H_1$}
\Text(70,25)[l]{$H_1$}
\Text(37,25)[r]{$H_2$}
\Text(37,55)[r]{$H_1$}
\Text(40,80)[c]{${\color{blue}Y^E_{\mu \mu}}$}
\Text(40,0)[c]{${\color{blue}\rho^E_{e \mu }}$}
\Text(37,39)[r]{${\color{blue}\Lambda_6}$}
\end{picture}
\qquad \qquad \qquad
\begin{picture}(60,100)(0,0)
\ArrowLine(0,0)(40,10)
\ArrowLine(40,10)(80,0)
\ArrowLine(0,90)(40,70)
\ArrowLine(40,70)(80,90)
\DashArrowLine(40,70)(40,40){1}
\DashArrowLine(40,40)(40,10){4}
\DashArrowLine(40,40)(70,60){1}
\DashArrowLine(70,20)(40,40){1}
\Text(10,-5)[r]{$\mu_R^-$}
\Text(72,-5)[l]{$e_L^-$}
\Text(10,100)[r]{$t_R$}
\Text(72,100)[l]{$t_L$}
\Text(70,55)[l]{$H_1$}
\Text(70,25)[l]{$H_1$}
\Text(37,25)[r]{$H_2$}
\Text(37,55)[r]{$H_1$}
\Text(40,80)[c]{${\color{blue}Y^U_{tt}}$}
\Text(40,0)[c]{${\color{blue}\rho^E_{e \mu }}$}
\Text(37,39)[r]{${\color{blue}\Lambda_6}$}
\end{picture}
\qquad\qquad\qquad
\begin{picture}(60,100)(0,0)
\ArrowLine(0,0)(40,10)
\ArrowLine(40,10)(80,0)
\DashArrowLine(40,70)(40,40){4}
\DashArrowLine(40,40)(40,10){4}
\Photon(40,40)(70,60){2}{4}
\Photon(40,40)(70,20){2}{4}
\DashArrowLine(40,95)(40,70){1}
\DashArrowLine(65,70)(40,70){1}
\DashArrowLine(40,70)(15,70){1}
\Text(10,-5)[r]{$\mu_R^-$}
\Text(72,-5)[l]{$e_L^-$}
\Text(70,55)[l]{$W$}
\Text(70,25)[l]{$W$}
\Text(37,25)[r]{$H_2$}
\Text(37,55)[r]{$H_2$}
\Text(15,70)[r]{$H_1$}
\Text(77,70)[r]{$H_1$}
\Text(37,90)[r]{$H_1$}
\Text(40,0)[c]{${\color{blue}\rho^E_{e \mu }}$}
\Text(45,80)[l]{${\color{blue}\Lambda_6}$}
\end{picture}
\end{center}
\end{boldmath}
\vspace{10mm}
\caption{
The left pair of diagrams generate the  
dimension eight $(\overline{L}_\mu H_1 E_\mu) 
(\overline{L}_e H_1  E_\mu)$ operator, by
 matching-out  the  heavy doublet Higgs $H_2$ (dashed line).
The third diagram  generates a dimension eight relative
of the  $(\overline{Q}_3  U_t) 
(\overline{L}_e  E_\mu)$ operator, see eqn (\ref{td8}).
The last diagram  gives a  $e \mu WW HHH $ interaction  by
 matching-out a heavy doublet Higgs $H_2$.
\label{fig:m1l} }
\end{figure}

All the terms in eqns  (\ref{1lh}) and (\ref{1ldc}), 
 are log-enhanced, so should arise
in the one-loop RGEs, however most of the terms are ${\cal O}(1/M^4)$.
This is because they arise in matching onto the 
 operator ${ O}^{e \mu  \mu \mu}_{S}$, which
would be of dimension eight if SU(2) was imposed:
$v^2 { O}^{e \mu  \mu \mu}_{S} = (\overline{L}_e H E_\mu) 
 (\overline{L}_\mu H  E_\mu)$.
For instance, in matching at the scale $M$, the two diagrams illustrated
on the left in figure \ref{fig:m1l}
will  contribute to the coefficient of the dimension eight operator:
\beq
\frac{~^{(8)}C^{e \mu  \mu \mu}}{ M^4} \sim  \frac{[\rho^E]_{e \mu } \Lambda_5[\rho^E]_{\mu \mu}}{ M^4}
~~~,~~~
  \frac{[\rho^E]_{\mu e} \Lambda_6[Y^E]_{\mu \mu}}{ M^4}
\eeq
The operator ${ O}^{e \mu  \mu \mu}_{S}$
can then mix via a penguin to the dipole operator.

Then there is also a light-higgs contribution to the
same operator at ${\cal O}(1/M^4)$, which can nonetheless
by relatively enhanced by $\tan \beta$ with respect to the 
 ${\cal O}(1/M^2)$ term. The corresponding diagram contracts
two of the diagrams illustrated on the right in figure \ref{fig:2},
so has four external $H_1$ lines.

\subsubsection{The top loop }

The two-loop contribution of $H$
 and $A$ is log$^2$ enhanced, so arises
 in the  one-loop RGEs.  In addition
to the  previously discussed dimension six term,
 that can be 
suppressed by $\cot \beta$, 
 the third diagram  of figure \ref{fig:m1l}
matches onto 
either  of two dimension eight operators 
$H_1^\dagger H_1 (\overline{L}_e^A  E_\mu )\epsilon_{AB} (\overline{Q}^B_t U_t )$
and 
$ (\overline{L}_e H_1  E_\mu ) (\overline{Q}_t \widetilde{H_1} U_t )$.
The second arises when  the $H_1$ contracted
with $H_2$ is an external leg. In both
cases, the coefficient is of order
\beq
\frac{^{(8)}C^{e \mu tt} }{ M^4} \sim 
\frac{ [\rho^{E }]_{e \mu} \Lambda_6[Y^{U}]_{tt}}{ M^4}
\label{td8}
\eeq
and such a term appears in the  full-model calculation
of eqn (\ref{fonctionnett}).

The top-loop contribution of  the light $h$,
which can contribute a significant part of
the $\meg$ amplitude,  
is a two-loop matching contribution at 
the weak scale.

\subsubsection{The $b$ loop }

The  $b$ loop  differs from the top loop,
 in that the heavy Higgs only contribute
at ${\cal O}(1/M^4)$, because there
is no dimension-six scalar operator
for $b$s that can mix to a tensor.  Matching   out
the heavy Higgs  onto the
dimension eight operator
$(\overline{L}_\mu H_1 E_e)  (\overline{Q}_3 H_1  D_b)$,
would give 
\beq
\frac{^{(8)} C^{e \mu  bb} } {M^4}\sim  \frac{[\rho^E]_{e \mu } \Lambda_5[\rho^D]_{bb}}{M^4}
~~~,~~~
\frac{[\rho^E]_{\mu e} \Lambda_6[Y^D]_{bb}} {M^4}
\eeq
then between  the scales $M$ and $m_b$,
the RGEs of QED mix the dimension-eight scalar operator 
$(\overline{L}_e H_1 E_\mu ) (\overline{Q}_b H_1 D_b )$
 to the dimension-eight tensor 
$(\overline{L}_e H_1 \sigma^{\mu\nu} E_\mu ) (\overline{Q}_b H_1 \sigma_{\mu\nu} D_b )$
 and the tensor to the  dimension eight dipole,
which reproduces that  heavy Higgs part of eqn (\ref{fonctionnebb}).

As in the one-loop contribution, there is an ${\cal O}(1/M^4)$
term  in the light-higgs exchange amplitude, which can be
tan$\beta$ enhanced, and arises due to two appearances
of the dimension six ${ O}_{eH}$, with indices $e \mu$
and $bb$.

\subsubsection{The W loop}

Despite that the $W$-loop
can give the dominant contribution
to $\meg$ in the 2HDM, none of it
was obtained in a one-loop EFT calculation
using dimension six operators. 

The heavy Higgs part
   has 
terms of  
 ${\cal O}(\log/M^2)$,
 ${\cal O}(\log^2/M^4)$,
and  ${\cal O}(\log/M^4)$. Consider here the first two:
CHK refer to the dimension six part as a ``non-decoupling''
contribution, because they take the mixing angle
$\cbma$ as a free parameter, rather than using the
decoupling limit dependence given in eqn (\ref{cbmadl}).
They say this contribution  arises from goldstone
loops, so in EFT it could  be generated  by matching
out the heavy doublet onto ${ O}^{e \mu}_{eH } $,
following by the mixing  of ${ O}_{eH}^{e \mu}$ to the dipole operator
in 2-loop RGEs.

Consider now the  ${\cal O}(\log^2/M^4)$ terms,
which  should  arise at second order in  the 1-loop
RGEs of dimension eight operators.
These could be generated by tree matching onto 
$(D_\mu H_1)^\dagger( D^\mu H_1)  \overline{L} H_1  E$
 as in the last diagram of  figure \ref{fig:m1l} (notice this operator
is symmetric on interchange of the Lorentz indices of the $W$),
 then mixing in the one-loop RGEs
of QED  to  a tensor operator such as 
$[(D_\mu H_1)^\dagger( D_\nu H_1) -
(D_\nu H_1)^\dagger( D_\mu H_1) ] (\overline{L} H_1 \sigma^{\mu \nu} E)$,
which  could then mix to a dimension eight
dipole $
( H_1^\dagger H_1)
( \overline{L} H_1 \tau^a \sigma^{\mu \nu} E)  W_{a, \mu \nu}$.

Finally, there is the light higgs contribution, which is
 ${\cal O}(1/M^2)$, and arises in two-loop matching at
the weak scale.

\end{document}
\bye